\documentclass[%
 reprint,
 amsmath,amssymb,
 aps,
pra,
]{revtex4-2}

\usepackage{graphicx}
\usepackage{dcolumn}
\usepackage[format=plain]{caption}

\usepackage{bm}
\usepackage{hyperref}
\usepackage[english]{babel}

\usepackage[table]{xcolor}
\usepackage{subcaption}
\usepackage{physics}
\usepackage{overpic}
\usepackage{booktabs} 
\usepackage{adjustbox} 
\usepackage{float}
\usepackage{tabularx}
\usepackage{pifont}
\definecolor{HeaderBlue}{HTML}{005D8F}
\definecolor{RowGray}{HTML}{F0F0F0}
\definecolor{CheckGreen}{HTML}{008000}
\definecolor{CrossRed}{HTML}{D2042D}
\newcommand{\cmark}{\textcolor{CheckGreen}{\ding{51}}}
\newcommand{\xmark}{\textcolor{CrossRed}{\ding{55}}}
\newcommand{\header}[1]{\rotatebox{90}{\parbox{4.5cm}{\centering\bfseries\color{white}#1}}}
\DeclareMathOperator{\sinc}{sinc}
\usepackage{ragged2e}
\begin{document}

\title{Making non-Markovian master equations accessible with approximate environments}

\author{Gerardo Su\'arez}
\email{gerardo.suarez@phdstud.ug.edu.pl}
\affiliation{%
 International Centre for Theory of Quantum Technologies, University of Gdansk, 80-308 Gdansk, Poland\\
}%
\author{Micha\l~Horodecki}
\affiliation{%
 International Centre for Theory of Quantum Technologies, University of Gdansk, 80-308 Gdansk, Poland\\
}
\begin{abstract}
Accurate and efficient simulation of open quantum systems remains a significant challenge, particularly for non-Markovian dynamics. We demonstrate the profound utility of expressing the environmental correlation function as a sum of damped sinusoidals within master equations. While not strictly required, this decomposition offers substantial benefits, crucially reducing the cost of Lamb-shift and decay rates calculations without sacrificing accuracy. Furthermore, this approach enables straightforward calculation of Lamb-shift corrections, bypassing the need for complex principal value integration. We show that these Lamb-shift effects are demonstrably non-negligible in heat transport scenarios, and are needed for an accurate description. Unlike in the Gorini-Kossakowski-Lindblad-Sudarshan (GKLS) master equation, the non-commuting nature of the Lamb-shift with the Hamiltonian in non-Markovian descriptions, coupled with GKLS's inaccuracies at early times, brings the necessity of non-Markovian descriptions for finite-time thermodynamics. In the weak coupling regime, our Master Equation formulations with exponential decomposition achieve accuracy comparable to numerically exact methods. This methodology significantly simplifies and accelerates the simulation of non-Markovian dynamics in open quantum systems, offering a more reliable and computationally tractable alternative akin to a Global Master Equation.
\end{abstract}

\maketitle

\section{Introduction}
Studies in the field of open quantum systems and quantum thermodynamics often 
employ a Gorini-Kossakowski-
Lindblad-Sudarshan (GKLS) master equation \cite{breuer,rivas}, however, that simple 
description has several known limitations 
\cite{onam,Hofer_2017,Cattaneo_2019,Levy_2014,ss2,meanforcereview,meanforce}
in particular, it fails to describe accurate dynamics in the transient regime 
\cite{dynamics,hartman,intermediate}, and structured or unstructured spectral densities look 
the same at this level of dynamics, which is both an attractive feature and a 
limitation \cite{breuer,intermediate}. On the other hand, there are multiple 
numerically exact approaches like the Hierarchical Equations of Motion (HEOM)
\cite{Numerically2020,bofin}, pseudomodes \cite{garraway_pseudomodes,Lambert2019,dampf},
 Time-Evolving Density Matrix using Orthogonal Polynomials Algorithm (TEDOPA) \cite{Rosenbach_2016,dand}, among others.  However, the cost of numerically exact
solutions tend to be large, which limits the kind of systems we can simulate effectively.
Most numerical methods approximate the environment  by decomposing it into pieces
that are easy to deal with, drawing inspiration from these methods, we aim to apply 
the same kind of approximation to non-Markovian master equations, descriptions that 
are between the standard GKLS master equations and 
numerically exact descriptions. 
\begin{figure}[H]    \includegraphics[width=0.85\linewidth]{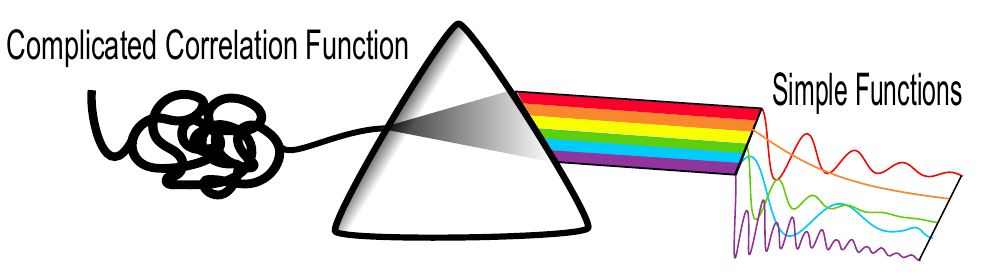}
    \caption{A cartoon illustrating the idea behind the decomposition of the correlation function} 
    \label{fig:illustration}
\end{figure}
Non-Markovian master equations are a terrific tool for the study of open quantum
systems, while their range of application is in principle limited, they can even 
lead to the exact solution of the system in some cases \cite{dynamics,Doll_2008},
they offer better descriptions than the GKLS master equation in the transient 
regime \cite{dynamics,hartman} and in some cases, their superiority extends to the 
steady state \cite{meanforcereview,ss2,meanforce}. Unlike in the case of a 
Lindblad master equation, non-Markovian dynamics generators usually involve integrals 
that need to be performed numerically, which slows down their simulation, furthermore 
the number of said integrals grows with the number of jump operators needed, thus 
the calculation of decay rates can be time consuming for this kind of method.

Futhermore, in the discussion of open quantum systems the Lamb-shift correction is often ignored,
in the last years, people have been advocating for its importance and inclusion 
\cite{winczewski2023renormalizationtheoryopenquantum,correa2024potentialrenormalisationlambshift,meanforcereview}
while in Markovian scenarios, the Lamb-Shift commutes with the system Hamiltonian and 
just induces an energy shift. In non-Markovian scenarios, it might not commute with 
the Hamiltonian of the system and it may induce non-trivial effects. We will show 
how the inclusion of Lamb-shift is needed to correctly describe heat currents using master
equations by using an example.

The Lamb-shift correction is often difficult to calculate because of the principal value in the inner integrals, making it computationally expensive. In this work, we demonstrate that our approximation for quantum environments allows the Lamb-shift to be determined through simple algebraic expressions. This approach provides a twofold advantage: it accelerates simulations and facilitates the inclusion of the Lamb-shift correction in a wider range of studies. We expect this to have far-reaching consequences in finite-time thermodynamics
as the inadequacy of the GKLS master equation to describe transient heat currents is 
expected to have effects on the optimal work, heat, and the limits of irreversibility, a topic that has been gaining 
popularity \cite{finitetime1,finitetime2,finitetime3,finitetime4,finitetime5,finitetime6,finitetime7,finitetime8,finitetime9} and 
has largely used local GKLS master equations.
We expect this to have a big effect whenever multiple parties make up the working body of the 
heat engine.

The way the approximation for the environment works is that we decompose
our original correlation function $C(t)$ into simpler functions that allow us to compute 
the required integrals analytically, figure  \ref{fig:illustration} shows an illustration 
of the idea. In this article drawing inspiration from numerically exact methods
\cite{Numerically2020,bofin,menczel,ishizaki_fmo,garraway_pseudomodes,Lambert2019} we 
decompose our function in terms of decaying exponentials, while in numerically exact methods, the Hilbert space to study grows with the number of exponents, here it remains constant. Another advantage of choosing 
this decomposition is that there are many methods to perform it in the literature \cite{takahashiHighAccuracyExponential2024}
, however,  other decompositions in terms 
of special functions or orthogonal polynomials are possible and worth exploring in the future.

The paper is organized as follows: in the first part, we review the two-point 
correlation function for thermal environments, we then show how to approximate it, 
the typical problems with the approximation and how they are overcome when one uses master 
equations, we then review and discuss the two main approaches used  in this article, 
the time convolutionless equation ($TCL$) in the second order ( $TCL_{2}$ aka Redfield equation) and the cumulant equation (aka refined weak coupling limit). Later, we provide four examples that prove the usefulness of our method, namely

\begin{itemize}
    \itemsep -0.2em 
    \item Example 1: The Spin-Boson Model. This example shows how accurate non-Markovian master equations can be, and it highlights that no accuracy is lost when using decaying exponential approximations for the correlation function.
    \item Example 2: A heat transport scenario. This example shows that non-Markovian descriptions and the Lamb-shift are necessary to accurately describe heat currents. This is no longer a ``textbook" example, as it illustrates the local vs. global debate and the cumulant method for multiple environments with interacting subsystems. Interaction is crucial here; as the fact that this equation tends toward the Global master equation as $t \to \infty$ is precisely what makes it ``worse" than the Redfield approach in \cite{hartman}, as it tends to the ``wrong steady state".
    \item  Example 3: A Kerr nonlinearity. This example shows that even for long times where the observables have reached a steady state, the refined weak coupling limit might be a better description than GKLS, contrary to expectations \cite{rwc,intermediate,dynamics}. We believe the slow convergence towards GKLS makes this example interesting, as it shows the cumulant equation might have a broader applicability than previously thought.  
    \item Example 4: Spin boson in a structured spectral density. This example showcases the use of exponential fitting in Volterra equations \cite{Gao2022,VABISHCHEVICH2022177} (which is more similar to the usage in HEOM or pseudomodes) and higher orders of $TCL$. It also shows how the methods described here can handle experimental spectral densities which are typically structured \cite{RatsepJL2007,lorenzoni}.
\end{itemize}
The code to reproduce the examples can be found in \cite{repo}. In this article, we will deal with bosonic environments only; however, the same 
scheme can be used for fermionic environments as well.  Throughout the manuscript, we use natural units, namely $\hbar = k_{B}=1$  and all reported times are measured in seconds.

\section{\label{sec:corr} Two point correlation function as decaying exponentials}
The interaction between a system and an environment can be fully determined by 
the system Hamiltonian $H_{S}$, the interaction operator $Q$, and the 
environment's correlation function in the case of Gaussian environments 
\cite{breuer} (or using either the power spectrum or the spectral density, and temperature). This quantity fully determines the dynamics of open quantum systems, 
however, we do not only require the correlation function but also some of its 
derivatives and integrals \cite{breuer,rivas}. Multidimensional integrals of 
the correlation function are often involved and can be prohibitively 
expensive. Often, such high-dimensional integrals are approximated to 
expressions valid only for the long time limit, to obtain a simpler 
Markovian description \cite{breuer,rivas}.
The correlation function for a thermal environment is given by 
\small
\begin{align}\label{eq:correlation_function}
   C(\tau) = \int_{0}^{\infty}\hspace{-1em} d\omega \frac{J(\omega)}{\pi} \left( \coth\left( \frac{\beta \omega}{2} \right) \cos \left(\omega \tau\right) - i \sin \left(\omega \tau\right)\right).
\end{align}
\normalsize
While its Fourier transform, the power spectrum is related to the spectral density via 
\small
\begin{align}
    S(\omega) &= \int_{-\infty}^{\infty}\hspace{-1em} dt e^{i \omega t} C(t) =  \operatorname{sign}(\omega)\, J(|\omega|) \Bigl[ \coth\Bigl( \frac{\beta\omega}{2} \Bigr) + 1 \Bigr].
\end{align}
\normalsize
The power spectrum obeys the detailed balanced condition.
\begin{align}
    S(\omega) = e^{\beta \omega} S(-\omega),
\end{align}
while in general, the correlation function cannot be obtained analytically, 
certain shapes for the spectral 
density that allows for a series representation of the 
correlation function \cite{Numerically2020,garraway_jc}. For example, 
using the Brownian motion underdamped spectral density
\begin{align}
    J_U(\omega) = \frac{\lambda^2 \Gamma \omega}{[(\omega_0^2 - \omega^2)^2 + \Gamma^2 \omega^2]}.
\end{align}
One can obtain an analytical answer for the correlation function in terms of an 
infinite sum of decaying exponentials.
\begin{figure}[H]
\includegraphics[width=\linewidth]{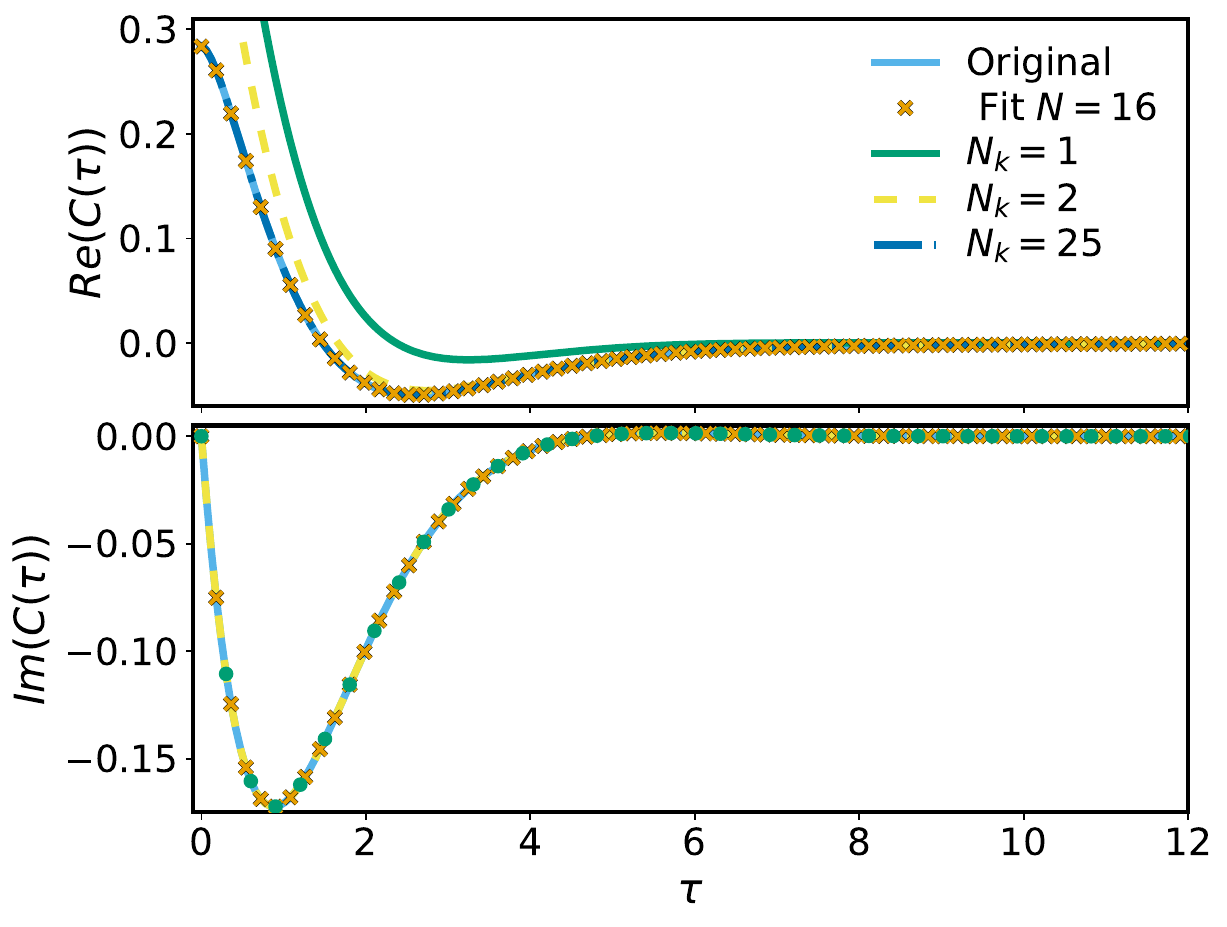}
    \caption{Real and imaginary part of the Correlation function of an underdamped Spectral density, the 
    convergence of the Matsubara expansion is shown, the parameters used for 
    the plot where  $\frac{\omega_{0}}{T}=24$, $\lambda^2= \frac{6\pi}{5} \omega_{0}^3$, $\Gamma = \frac{10\omega_0}{6}$. For an underdamped spectral density, only one exponent is 
    needed for the imaginary part of the correlation temperature, while the real part converges slowly.} 
    \label{fig:correlation}
\end{figure}
Using this sort of expansion for the correlation function, Kubo and Tanimura expressed 
the exact dynamics of an open quantum system coupled to a Gaussian environment
in terms of the Hierarchical Equations of Motion (HEOM) \cite{Numerically2020,Time} 
and since then many other methods require such a decomposition 
\cite{Lambert2019,menczel,smirne,taming}. Those methods usually employ the 
Matsubara decomposition \cite{Numerically2020,Lambert2019,ishizaki_fmo} 
to write the correlation function as a finite sum of decaying exponentials 
\cite{bofin}, however, there are other ways to obtain this representation \cite{takahashiHighAccuracyExponential2024,qutip5} (see  Appendix \ref{app:exponents} for a brief overview of popular methods), for example for the underdamped spectral density mentioned above
\begin{equation}\label{eq:exps}
    C(\tau) = \sum_{k=0}^{\infty} c_{k} e^{-\nu_{k} \tau},
\end{equation}
where 
\begin{equation}
    \nu_k = \begin{cases}
               -i\Omega  + \Gamma/2 & k=0 \\
               i\Omega  +\Gamma/2   & k = 1\\
               {2 \pi k T}   & k \geq 2,\\
           \end{cases}
\end{equation}
\begin{align}
        c_k &= \begin{cases}
               \left(\lambda^2 \coth(\beta( \Omega + i\Gamma/2)/2)+ i\alpha^2 \right)/4\Omega & k = 0\\
               \left(\lambda^2 \coth(\beta( \Omega - i\Gamma/2)/2) - i\alpha^2 \right)/4\Omega& k = 1\\
 \frac{-4\lambda^2\Gamma  \pi k}{\beta^{2}[(\Omega + i\Gamma/2)^2 + (\frac{2 \pi k}{\beta})^2)][(\Omega - i\Gamma/2)^2 + (\frac{2 \pi k}{\beta})^2]}     & k \geq 2\\
           \end{cases}
\end{align}
where $\Omega= \sqrt{\omega_{0}^{2}-\left(\frac{\Gamma}{2}\right)^{2}} $. The usefulness of this expansion (known as the Matsubara expansion \cite{Numerically2020,Mustafa_2023,ishizaki_fmo,bofin}) is largely due to two simple reasons.
\begin{figure}[H]
    \includegraphics[width=\linewidth]{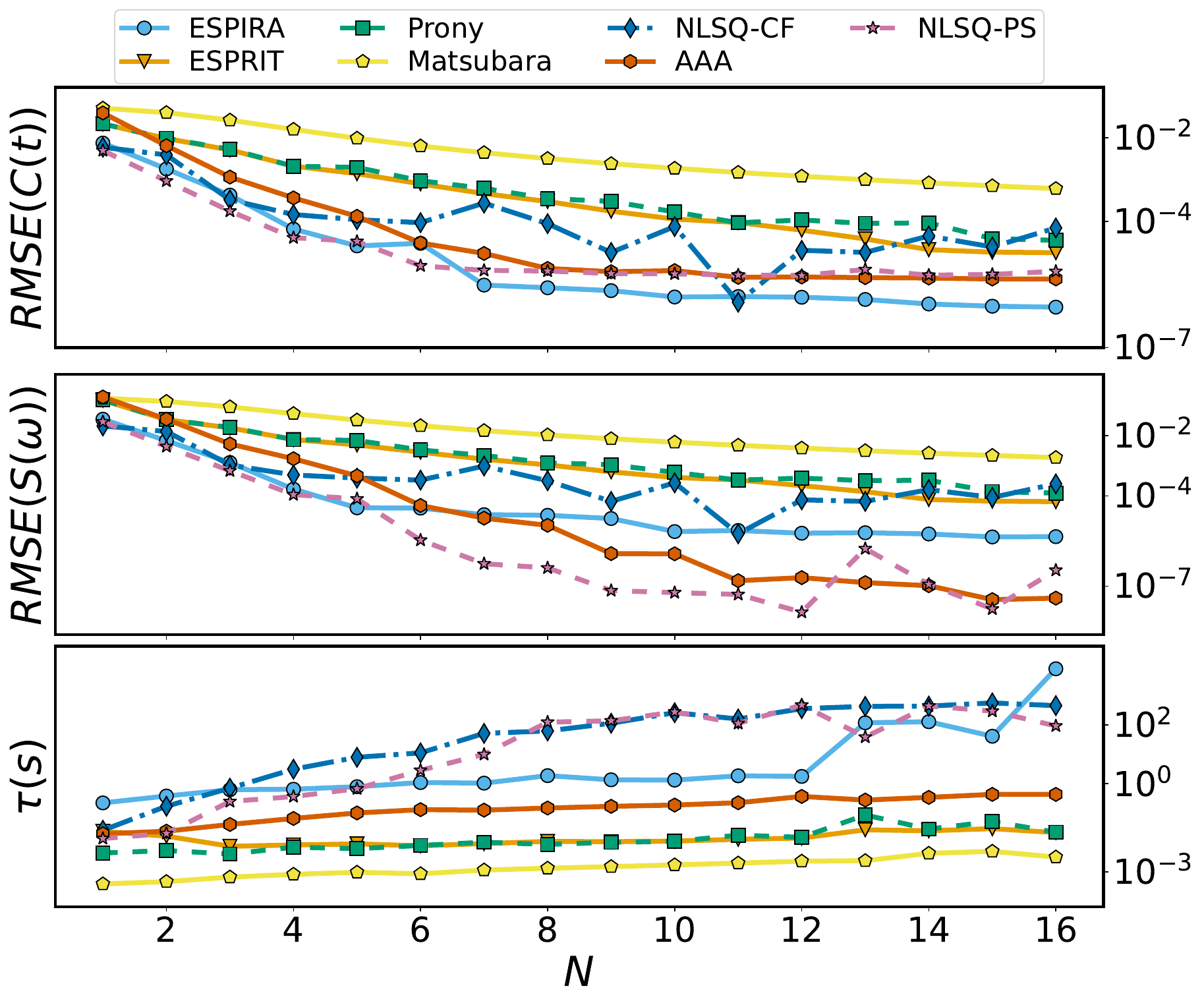}
    \caption{Shows the number of exponents needed to achieve a certain root mean 
    squared error (RMSE). We show the error for several methods to obtain the exponents. The top
    image shows the RMSE on the correlation function, the middle one the RMSE on the power spectrum
    and the bottom one is the time it takes each method to obtain the exponents. The correlation function parameters are 
    the ones from Figure \ref{fig:correlation}} 
    \label{fig:correlation_methods}
\end{figure}
\begin{enumerate}
  \item The sum converges quickly when the $T \gg \frac{\omega_{0}}{2 \pi}$, giving us the ability to express the correlation function as a sum of a small number of decaying exponentials
  \item The correlation function expressed this way, has simple derivatives that allow for the derivation of useful dynamical equations such as HEOM
\end{enumerate}

while its  main drawbacks are
\begin{enumerate}
  \item In the regime where $T \to 0 $, the expansion converges slowly, needing many exponents to correctly describe the correlation function, as figure \ref{fig:correlation} shows.
  \item Only a handful of spectral densities allow for such a nice, analytical expansion.
  \item The Fourier transform of the correlation function (The power spectrum  $S(\omega)$) may not be correctly approximated, leading to violations of the detailed balance condition 
    \cite{Lambert2019,ancilla}, in the cumulant equation used in this manuscript, this may lead to the loss of complete positivity, depending on the technique employed to obtain the exponents. This
    can be solved by including more exponents or changing the support points used for the approximation (see Appendix \ref{app:exponents}).
\end{enumerate}
One of the disadvantages of methods that make use of this decomposition is that 
they typically enlarge the Hilbert space of the system significantly 
\cite{bofin,Numerically2020,smirne,Lambert2019,menczel}. 
Those methods emphasize approximating the environment with 
the least number of exponents possible, while still maintaining the detailed balanced condition,  which can be challenging in practice.
Many methods have been devised to obtain such approximations of the correlation 
function of arbitrary spectral densities (see \cite{takahashiHighAccuracyExponential2024} 
for a review, and the Appendix \ref{app:exponents} for an overview of the methods applied in this manuscript). 
As Figure \ref{fig:correlation_methods} shows for a fixed number of exponents, 
different methods achieve different accuracies \footnote{Which method works better depends 
largely on the temperature of the environment, and the sampling points used. In general,
we recommend ESPIRA or NLSQ for transient dynamics and AAA when one wants to obtain the
steady state}. Typically a $RMSE \approx 10^{-4}\mathrm{max}(|C(t)|)$ on the correlation function 
is enough to accurately capture the dynamics of the system with the methods used in
this manuscript, on the other hand, having a low error on the power spectrum at the 
Bohr frequencies of the system, helps us satisfy the detailed balance condition. Table \ref{tab:visual_comparison}
briefly summarizes the main advantages and disadvantages of the methods shown in  Figure \ref{fig:correlation_methods}.
We recommend using Estimation of Signal Parameters based
on Iterative Rational Approximation (ESPIRA), specifically algorithm 1 from \cite{esprit}, as we have found it often delivers the best results.
In this paper, we show the usefulness of these methods in master equations, 
methods that do not require this decomposition but that greatly benefit 
from it. We put particular emphasis on the cumulant equation (Refined Weak 
coupling limit)\cite{dynamics,rwc,intermediate} a method for non-Markovian dynamics
that preserves positivity.  We show that unlike the exact methods, 
where using many exponents is prohibitively expensive, as the Hilbert space studied grows 
with the number of exponents, in master equations, using many exponents is 
not an issue, as the Hilbert space remains constant, and the computational cost does not 
increase too much. Using this approximation also enables the calculation 
Lamb-shift corrections which are usually neglected due to the 
complexity of their calculation. We show that the Lamb-shift correction is 
non-negligible in heat transport scenarios. This decomposition may also serve 
to enhance the interpretability of semi-analytical solutions in the 
non-Markovian regime, like the ones in \cite{dynamics}, as it makes 
non-Markovian equations similar to a Global master equation, achieving 
with higher precision than the regularizations in \cite{intermediate}.
\section{ The Cumulant Equation}\label{sect:decays}
Computing the decay rates in non-Markovian master equations can be costly, while one may think it is not a limiting factor, the number of decay rates that need to be computed grows with the number of Bohr frequencies, and computing each decay rate involves an integral that can be costly,
especially in the long time limit where highly oscillatory functions are often
involved. In general, computing decay rates involves integration and its dimensionality varies with the level of accuracy 
of the master equation. In the case of the cumulant equation, one-dimensional 
integrals are used for the decay rates, and a two-dimensional integral with 
a principal value in the inner integral is used for the Lamb-shift correction.

In this section, we show how to approximate the decay rates cumulant equation 
(aka Refined weak coupling limit) \cite{dynamics,rwc,rivas_thermo,hartman,Kosloff}, using a series of 
decaying exponentials for the correlation function. We perform this approximation 
to obtain an accurate and fast simulation technique.  The cumulant 
 equation is determined by the exponential of the map:
\begin{align}
&\mathcal{K}_t [\rho(0)]= - i \sum_{\omega,\omega'}\xi(\omega,\omega',t) [ A^{\dagger}(\omega) A(\omega'),\rho(0)]  \\
&+\underset{\omega,\omega'}{\sum} \Gamma(\omega,\omega',t) \Big( A(\omega') \rho(0) A^{\dagger}(\omega) \nonumber  - \frac{1}{2} \{ A^{\dagger}(\omega) A(\omega'),\rho(0)\} \Big),
\end{align}
where
\footnotesize
\begin{align}
\Gamma(\omega,\omega',t) &=\int_{0}^{t} dt_1 \int_{0}^{t} dt_2 e^{i (\omega t_{1} - \omega' t_{2})} C(t_{1}-t_{2}), \\
\xi(\omega,\omega',t) &= \iint_{0}^{t} \frac{dt_{1} dt_{2}}{2i} \text{sgn}(t_{1}-t_{2}) e^{i(\omega t_1- \omega' t_2)} C(t_{1}-t_{2}),
\end{align}\normalsize
and we used the standard jump operator representation
\begin{align}
    H_{I} &= \sum_{k} A_{k} \otimes B_{k}  = \sum_{k,\omega} e^{-i \omega t} A_{k}(\omega) \otimes B_{k}, \\ 
    A(\omega) &= \sum_{\epsilon-\epsilon' =\omega} \ketbra{\epsilon} A \ketbra{\epsilon'} .
\end{align}
The dynamics of the system are then given by:
\begin{equation}\label{eq:linear_map}
    \rho(t)=e^{\mathcal{K}_{t}} [\rho_S(0)],
\end{equation}
when we consider thermal environments, the time-dependent coefficients simplify to \footnote{There is a factor $\pi$ difference with respect to some articles \cite{rwc,dynamics} due to differences in the definition of the correlation function Eq. \eqref{eq:correlation_function}}
\begin{align} 
&\Gamma(\omega,\omega',t) = \frac{t^{2}}{\pi}\int_{0}^{\infty} d\nu e^{i\frac{\omega-\omega'}{2} t} J(\nu) \left[ (n(\nu)+1) \sinc\left(\frac{(\omega-\nu)t}{2}\right) \right. \nonumber \\ &\times \left. \sinc\left(\frac{(\omega'-\nu)t}{2}\right)   + n(\nu) \sinc\left(\frac{(\omega+\nu)t}{2}\right) \sinc\left(\frac{(\omega'+\nu)t}{2}\right)   \right], \label{eq:cum_decay}\\
&\xi(\omega,\omega',t)=\frac{t^{2}}{2 \pi^{2}} \int_{-\infty}^{\infty} d\phi \sinc\left( \frac{\omega-\phi}{2} t\right) \sinc\left( \frac{\omega'-\phi}{2} t\right) \nonumber \\ &\times P.V \int_{0}^{\infty} d\nu J(\nu) \left[ \frac{n(\nu)+1}{\phi-\nu} + \frac{n(\nu)}{\phi+\nu} \right],\label{eq:cum_lam}
\end{align}
where $n(\nu)$ denotes the Bose-Einstein distribution.  By representing the correlation function as a series of damped sinusoidals, in Appendix  \ref{app:cumulant},  we show that we can obtain these expressions algebraically:
\begin{align}
\chi_{k}(\omega,\omega',t) &= \frac{c_{k} e^{-(\nu_{k}-i\omega)t}}{(\nu_{k}-i \omega)(\nu_{k}-i \omega')} \nonumber
 \\&+ \frac{i c_{k}}{\omega-\omega'} \left( \frac{1}{(\nu_{k}-i \omega)} - \frac{e^{i(\omega-\omega') t}}{(\nu_{k}-i \omega')} \right),\\
\Gamma(\omega,\omega',t) &= \sum_{k}  \overline{\chi_{k}}(\omega',\omega,t) + \chi_{k}(\omega,\omega',t)  \\
\xi(\omega,\omega',t) &= \sum_{k} \frac{\chi_{k}(\omega,\omega',t) -  \overline{\chi_{k}}(\omega',\omega,t)  }{2 i}.
\end{align}

\begin{figure}[H]
\begin{overpic}[width=\linewidth]{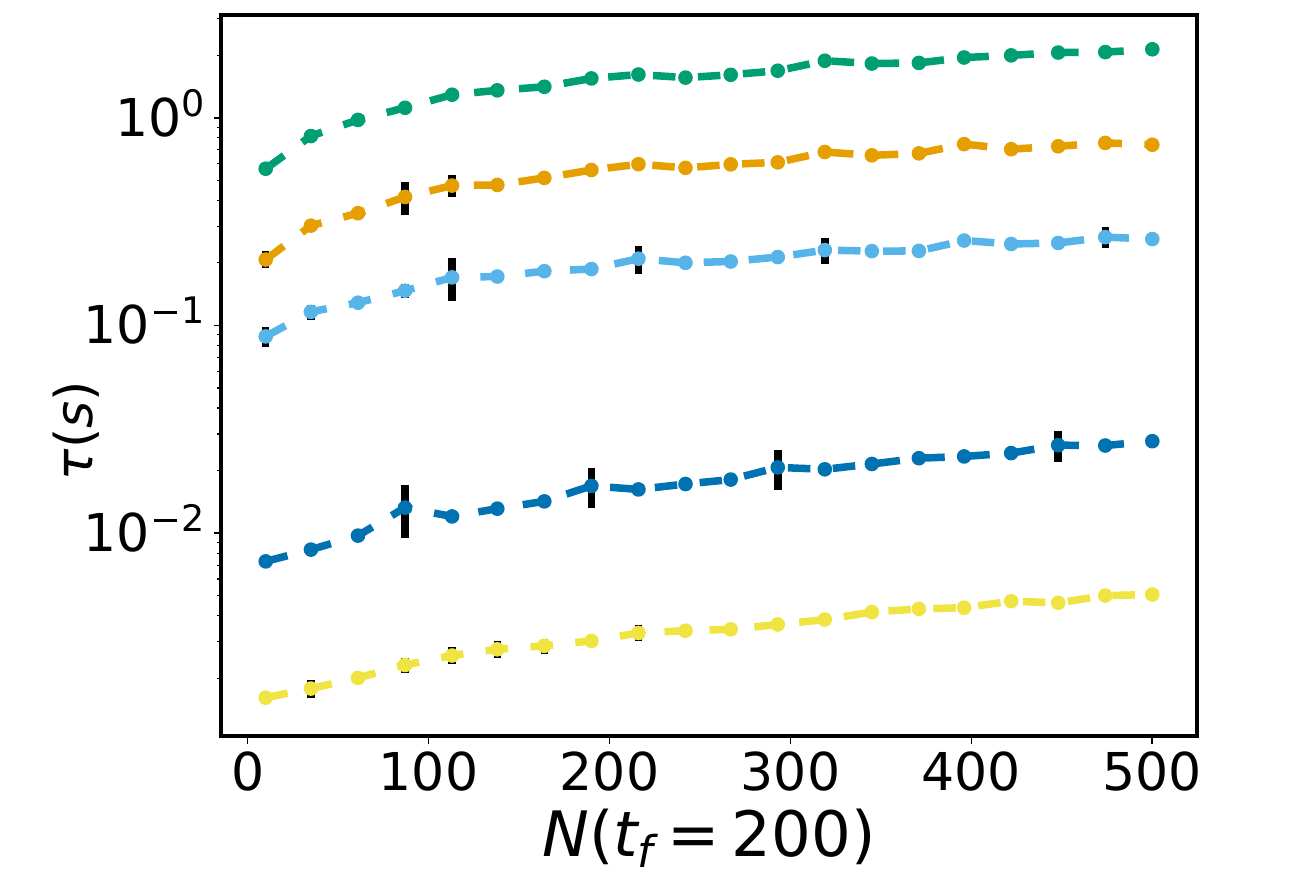}
    \put (3,65) {\large$\displaystyle (a)$}
\end{overpic}
\begin{overpic}[width=\linewidth]{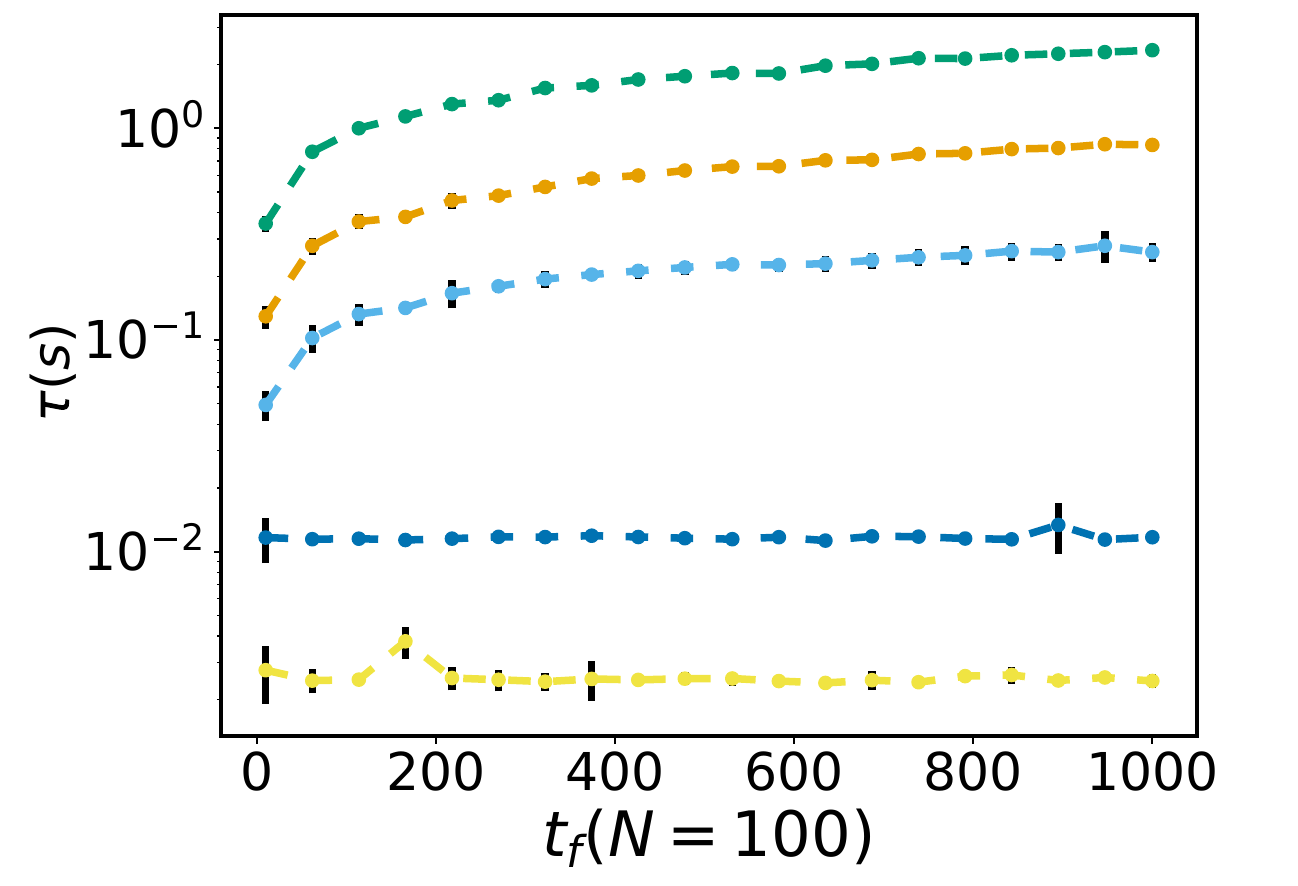}
    \put (3,65) {\large$\displaystyle (b)$}
\end{overpic}
\begin{overpic}[width=\linewidth]{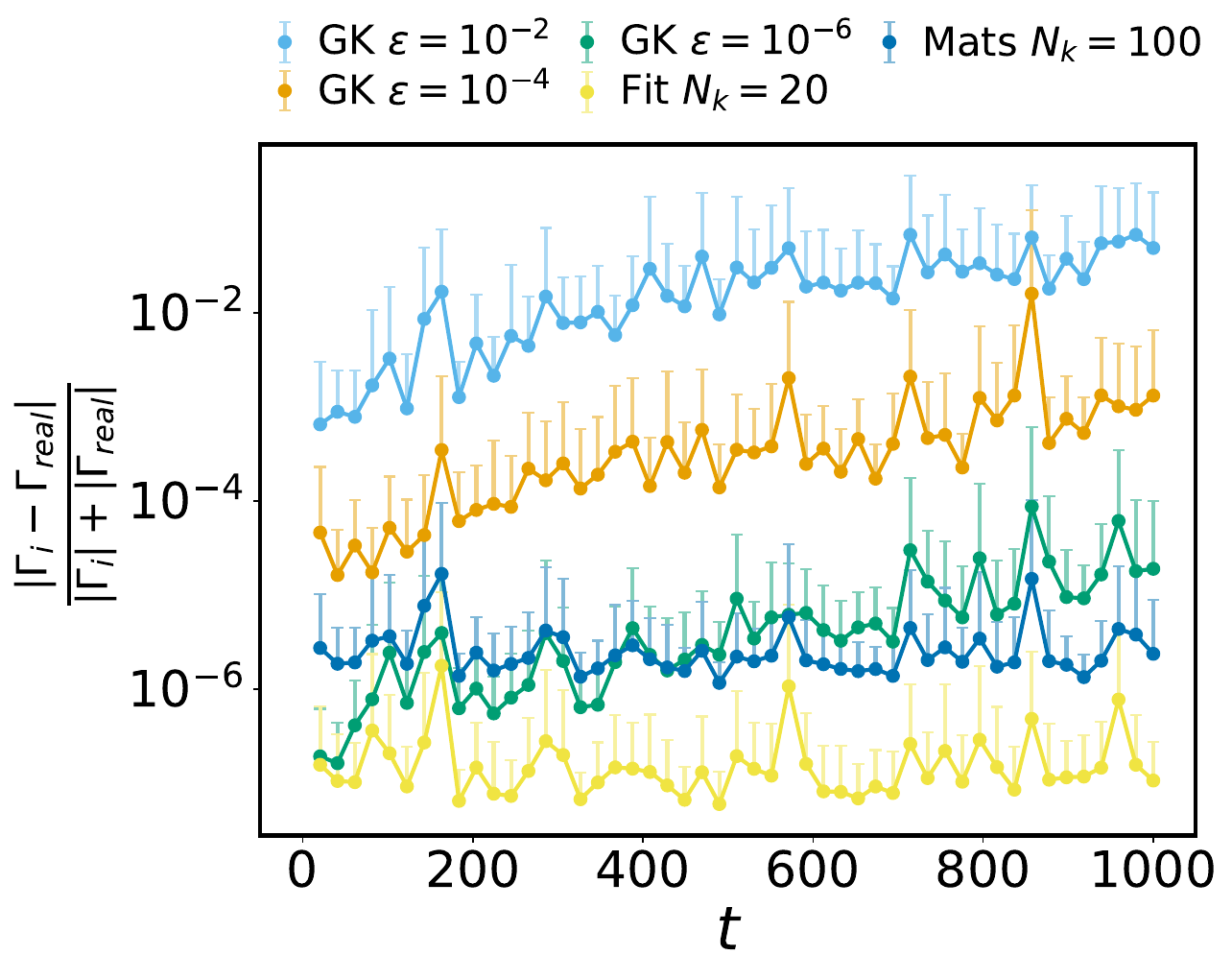}
        \put (0,65) {\large$\displaystyle (c)$}
\end{overpic}

    \caption{The plots show how the approximation of the decay rates affects
its computation time and precision, In these plots, ``GK" stands for 
the Gauss-Kronrod method for integration with 21 points, and $\epsilon$ 
is the tolerance required during the numerical integration procedure. 
These plots are aggregated from 50 randomly chosen $\omega,\omega'$.  In all panels, vertical bars indicate the standard deviation. 
a) The time it takes to compute the decay rate with a fixed final 
time $t_{f}=200$ using N time steps, we see for this particular 
example one gets a speed up of around 3 orders of magnitude.
b)Time taken to obtain decay rates as time increases with $N=100$ points,
c) Average relative error of the approximation of the decay rates. All the plots share the same legend.
}\label{fig:cum_time}
\end{figure}

The error of the approximation is bounded by (see Appendix \ref{app:bounds} for details)  
\begin{align}\label{eq:cum_bounds}
   &|\Gamma(\omega,\omega',t)-\Gamma_{\text{approx}}(\omega,\omega',t)|     \leq \begin{cases}\frac{ 4 \epsilon_{S}  \pi }{|\omega-\omega'|}  \text{ for } \omega \neq \omega' \\ 
        2\pi \epsilon_{S} t \end{cases} \\ 
   &|\xi(\omega,\omega',t)-\xi_{\text{approx}}(\omega,\omega',t)|     \leq \begin{cases}\frac{ 2 \epsilon_{S}  \pi }{|\omega-\omega'|}  \text{ for } \omega \neq \omega' \\ 
        \pi \epsilon_{S} t \end{cases}
\end{align}
where $\epsilon_{S}$ is the absolute error on the power spectrum, which we assume to be constant. In Figure \ref{fig:cum_time}c, 
we see that the errors in approximating integrals are similar to the errors 
 $\epsilon_S$ (error in the power spectrum) obtained by approximating the correlation function by exponents (approximations using a fit for the exponents $\epsilon_{S}\approx 7.9 \times 10^{-7}$  and the Matsubara expansion $\epsilon_{S}\approx 1.8 \times 10^{-6}$).  They are slightly fluctuating around $\epsilon_S$
 since we averaged over random $\omega,\omega'$ in the range from $[0,1]$. Notice, the error in Figure \ref{fig:cum_time} does not grow with $t$ as one would expect from the bound \eqref{eq:cum_bounds}, the reason for this is that the magnitude of the integral itself grows in $t$ as one could tell from Eqs.\eqref{eq:cum_decay} and \eqref{eq:cum_lam} and thus the relative error is roughly constant in time (see Appendix \ref{app:bounds} for more details). The algebraic expressions in this section free us from the burden of numerical integration to obtain these coefficients, which, as we will see in the next sections, have a big impact on the total simulation time.

\section{The second order time convolutionless equation ($TCL_{2}$)}
The Time convolutionless at second order $TCL_{2}$ (aka the Redfield equation \footnote{In the literature, often the Redfield equation name refers to the one with decay rates computed with $t\to\infty$, here we refer to the one with time dependent coefficients as Redfield and the one with time independent coefficients as Bloch-Redfield.}) is a powerful master equation whose main drawback is its lack of complete positivity. As shown in \cite{hartman}, this is often not a limitation when working in its regime of validity. It can be written as \cite{meanforce,dynamics,breuer}: 
\begin{align} \label{eq:redfield}
&\frac{d}{dt} \rho(t) =  i \sum_{\omega,\omega'}  \tilde{\mathcal{S}}(\omega,\omega',t) [ \rho(t),  A^\dag (\omega) A (\omega')]\\&+  \sum_{\omega,\omega'}   \tilde \gamma(\omega, \omega', t) \left( A (\omega') \rho(t) A^\dag(\omega) -\frac{1}{2} \{A^\dag (\omega) A (\omega'), \rho(t) \} \right) , \nonumber 
\end{align}
where
\begin{align}\label{eq:redfield_coeffs}
\tilde S(\omega,\omega',t)&= \frac{1}{2 i}e^{i(\omega-\omega')t}(\Gamma(\omega',t)-\overline{\Gamma(\omega,t)}), \\
 \tilde \gamma(\omega,\omega',t)&= e^{i(\omega-\omega')t}(\Gamma(\omega',t)+\overline{\Gamma(\omega,t)}), \\
\Gamma(\omega,t)&=\int_{0}^{t}ds e^{i \omega s} C(s),
\end{align}
by substituting \eqref{eq:correlation_function} into \eqref{eq:redfield_coeffs} we obtain
\begin{align}
   \tilde S(\omega,\omega',t)&= \frac{1}{\pi}\int_{0}^{\infty} d\omega J(\omega) \left( A(\omega,\omega') +B(\omega,\omega')\right),
\end{align}
where
\small
\begin{align}
    A(\omega,\omega') &= \frac{2 i \omega'}{(\omega'-\omega)(\omega+\omega')} -i \left( \frac{e^{i t(\omega'-\omega)}}{\omega'-\omega} + \frac{e^{i t(\omega'+\omega)}}{\omega'+\omega} \right)
\end{align}
\begin{align}
    B(\omega,\omega') &= \frac{i(e^{it(\omega'-\omega)}-1)}{\omega-\omega'}. 
\end{align}

\begin{figure}[H]
    \centering
\begin{overpic}[width=\linewidth]{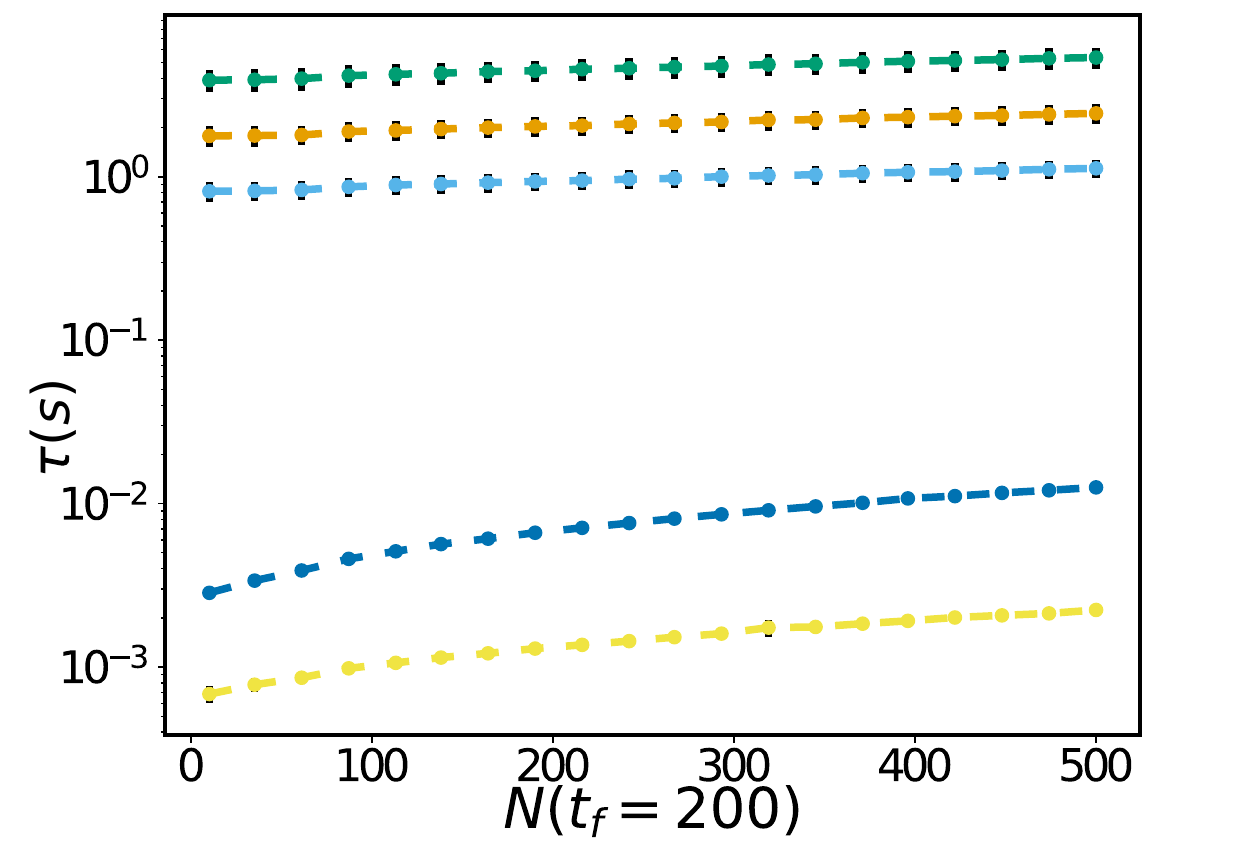}
    \put (3,65) {\large$\displaystyle (a)$}
\end{overpic}
\begin{overpic}[width=\linewidth]{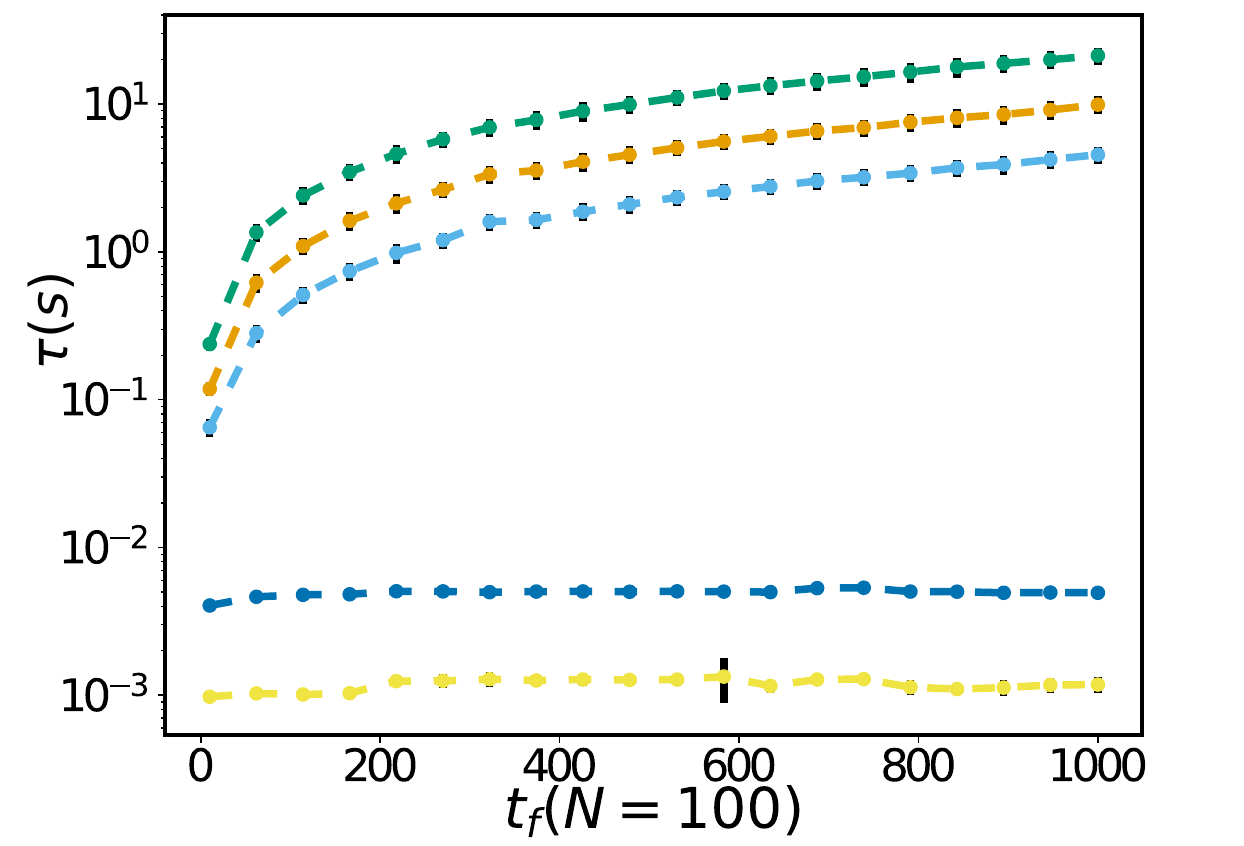}
    \put (3,65) {\large$\displaystyle (b)$}
\end{overpic}
\begin{overpic}[width=\linewidth]{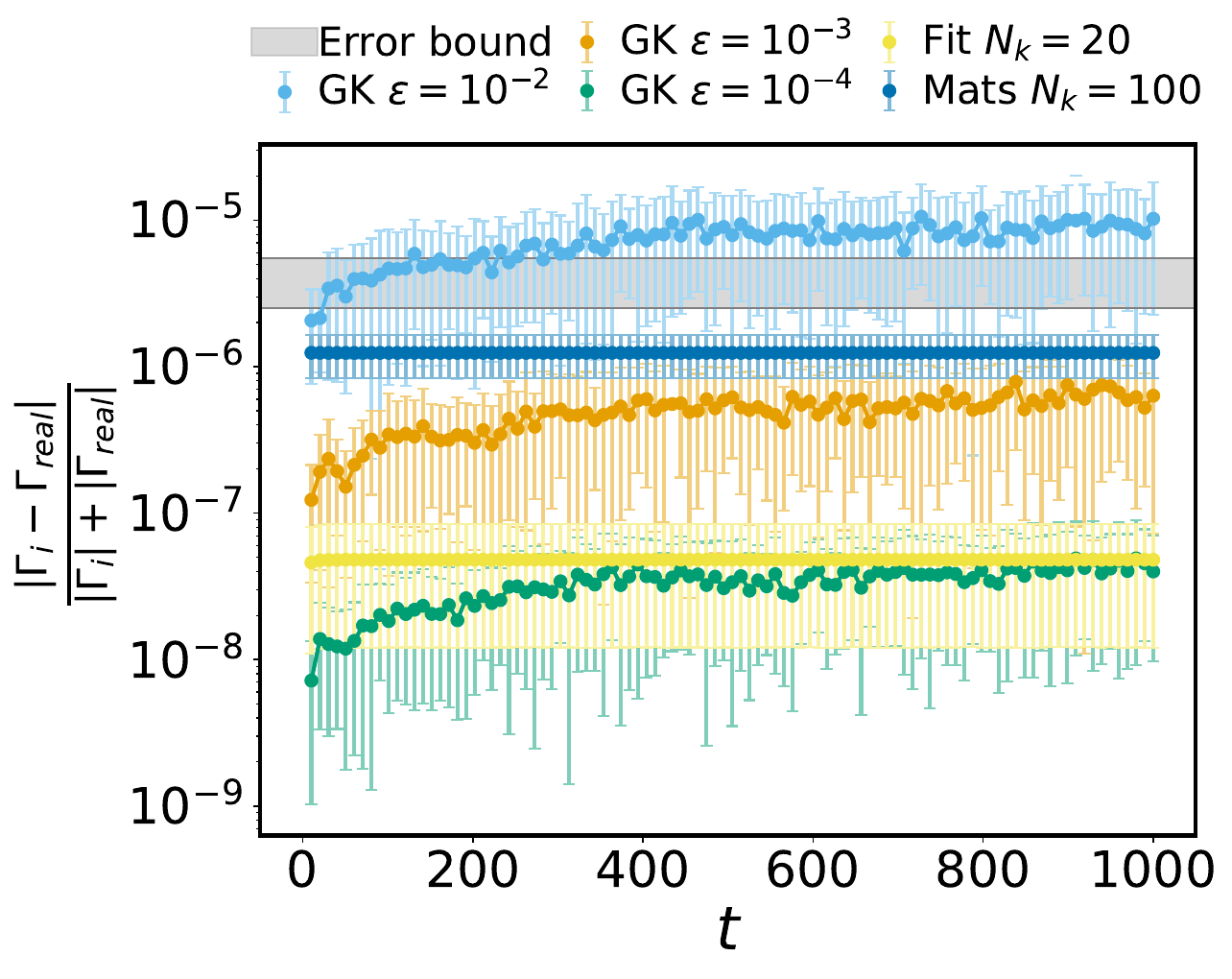}
        \put (0,65) {\large$\displaystyle (c)$}
\end{overpic}
    \caption{The plots show how the approximation of the decay rates affects
its computation time and precision. In these plots, ``GK" stands for 
the Gauss-Kronrod method for integration with 21 points, and $\epsilon$ 
is the tolerance required during the numerical integration procedure. 
These plots are aggregated from 50 randomly chosen $\omega,\omega'$.  In all panels, vertical bars indicate the standard deviation.  
a) The time it takes to compute the decay rate with a fixed final 
time $t_{f}=200$ using N time steps, we see for this particular 
example one gets a speed up of around 3 orders of magnitude.
b)Time taken to obtain decay rates as time increases with $N=100$ points,
c) Average error of the approximation of the decay rates, the shaded region indicates the error bounds, the top line is the Matsubara error bound while the lower corresponds to the fit one. All the plots share the same legend.
}\label{fig:time_red}
\end{figure}

\normalsize
By using the decaying exponential representation of the environment, one has 
\begin{align}
\Gamma(\omega,t)&= \sum_{k} c_{k}\int_{0}^{t}ds e^{-(\nu_{k}-i\omega)s}= \sum_{k} c_{k}\frac{1-e^{-(\nu_{k}-i\omega)t}}{(\nu_{k}-i\omega)}.
\end{align}
Those expressions allow us to obtain both the decay rate and the Lamb-Shift of the Redfield equation easily. In Fig. \ref{fig:time_red}, we see how it becomes advantageous to use the decaying exponential representation of the environment, as we obtain the decay rates quickly and accurately. The error of our approximation (see Appendix \ref{app:bounds})  is bounded by the error on the power spectrum ($\epsilon_{S}$) and is independent of the jump operator frequencies or time
\begin{align}
    |\gamma(\omega,\omega',t)-\gamma_{\text{approx}}(\omega,\omega',t)|  \leq 2 \pi \epsilon_{S}
\end{align} 
The validity of the bound is shown in Figure \ref{fig:time_red}c, where we see that both approximations yield constant error in time when averaged over random values for $\omega,\omega'$ and satisfy the bound. Using a fit for the exponents we have  $\epsilon_{S}\approx 7.9 \times 10^{-7}$  and the Matsubara expansion $\epsilon_{S}\approx 1.8 \times 10^{-6}$ 

The obtained speed-up is especially important for high-dimensional systems where the number of decay rates grows quickly with the number of transitions.
\section{Time convolutionless equation at higher orders $TCL_{2n}$}\label{sect:tcl_gen_approx}

In the previous section, we discussed the lowest non-trivial order for the time convolutionless equation in a thermal environment. If we consider higher orders, we can notice that
regardless of the order of the approximation, when considering thermal environments, the ordered cumulants will be composed of a combination of the system operators at different times, which we can write in terms of jump operators, and the product of $n$ correlation functions with different time arguments \cite{breuer,Lampert,smirnejc}, once we expand the ordered cumulants. The coefficients that accompany the operators in the generator will always have the form 
\begin{align} 
&\int_{0}^{t} dt_{1}\int_{0}^{t_{1}} dt_{2}  \dots\int_{0}^{t_{2n-2}} dt_{2n-1} C(t-t_{2})\dots C(t_{1}-t_{2n-1}) \nonumber \\
&\times e^{i\omega_{1}t\pm i\omega_{2}t_{2} \pm \dots\pm i \omega_{n-1}t_{1}\pm i\omega_{n}t_{n-1}}.
\end{align}
Although the coefficients are apparently complicated, notice that if we use Eq. \eqref{eq:exps} then this is just a high dimensional integral of an exponential function.
\begin{align} 
\sum_{\vec{k}} & \int_{0}^{t} dt_{1}\int_{0}^{t_{1}} dt_{2} \dots\int_{0}^{t_{2n-2}} dt_{2n-1} c_{k_{1}}e^{-\nu_{k_{1}}(t-t_{2})}\dots \nonumber \\
& c_{k_{n}} e^{-\nu_{k_{n}}(t_{1}-t_{n-1})}   \times e^{i\omega_{1}t\pm i\omega_{2}t_{2} \pm \dots\pm i \omega_{n-1}t_{1}\pm i\omega_{n}t_{n-1}} 
\end{align}
By doing the integral analytically, we will turn this high-dimensional integral into a multi-index sum. Notice that the decay rates in previous sections, were approximated by a sum of 
$m$ elements, where $m$ is the number of exponents we used to describe our two 
time correlation function, as the sum was over only one index is computational 
cost was $O(m)$. However, from the $TCL_{2n}$ expansion with ordered cumulants, 
we see that after approximation we will have $n$ indices in the summation 
(as it will be an integral of the product of $n$ correlation functions),
increasing the computational cost to  $O(m^{n})$, thus for higher order master equations, 
reducing the number of exponents used has a significant impact on the 
computational cost of the simulation.  This is still advantageous, typically an n-dimensional integral has a complexity of $~\mathcal{O}(\epsilon^{-\frac{d}{k}})$   where $d$ is the dimension and $k$ depends on the method used (for example, the trapezoidal rule has $k=2$), $\epsilon$ is the error tolerance. While, for Montecarlo integration the complexity goes as $\mathcal{O}(\epsilon^{-2})$  \cite{novak}. As shown in Figure \ref{fig:correlation_methods} a moderate number of exponents reaches accuracies of around $10^{-6}$. On the other hand, the error on the approximation of these integrals is bounded by (see Appendix \ref{app:bounds})
\begin{align}
    &||I_{n}(t) - I_{n}^{A}(t)|| \leq \left( \sum_{k}|c_{k}|\log \left(4\left(\frac{\omega_{c}}{\nu_{k}^{R}}\right)^{2}\right) \right)^{n-1} \nonumber \\ &\times  \left(\frac{\pi \epsilon_{S} t^{2n-2}}{(2n-2)!}  + \epsilon \frac{t^{2n-3}}{(2n-3)!}  (n-1) \right). 
\end{align}
where $\omega_{c}$ is a cutoff frequency for the environment such that  $S(|\omega| > \omega_{c})=0$. Just as in the case of the refined weak coupling (section \ref{sect:decays}), the bound growing in $t$ is not a problem as the integrals themselves grow with the same power of $t$ and the relative error stays roughly constant. Given that a relatively low number of exponents give small $\epsilon,\epsilon_{s}$, the $\mathcal{O}(m^{n})$ scaling still provides a low-cost alternative for the estimation of these integrals.

While the idea seems a bit abstract in this section, we will have a practical example in section \ref{sect:example4}.

\section{Examples}
In this section, we compare the use of the decaying exponential representation, 
to numerical integration and also a state-of-the-art numerically exact approach.
Let us note during these comparisons that we use the HEOM 
implementation from \cite{bofin}.
Bear in mind, that in this section no effort was made to optimize the Redfield
and cumulant equations beyond the approximation of decay rates, 
in particular for Redfield we use scipy's \textit{quad\_vec} and \textit{solve\_ivp}
\cite{2020SciPy} directly, which puts Redfield at a disadvantage
when the Hilbert space grows, therefore, the Redfield timing should be compared to itself
with or without our approximation of the decay rates. Similarly, the number of points used for the solution impacts the required time for computation more when using a dynamical map, than when solving the differential equation. The exponents in this 
section were obtained via Nonlinear squares fitting for the second example, in all other examples ESPIRA \cite{esprit} was used. The code to reproduce this section can be found in \cite{repo}. The Ohmic spectral density used in some of the examples is given by 
\begin{align}
    J(\omega) = \alpha\frac{\omega^s}{\omega_c^{s-1}} e^{-\omega / \omega_c} .
\end{align}
We would also like to remark that we used the same effective environment for HEOM and the master equations, and that significant effort was done to obtain a minimal number of exponents to improve HEOM's performance while not compromising accuracy, this optimization of the fitting is not necessary for the master equations as the impact of more exponents is minimal  (for more details see Appendix \ref{app:onemore} )

\subsection{Example I: The Spin Boson Model}\label{sect:example1}
In this section, we consider the  Spin-Boson Model, a comparison of the methods 
used in this paper when applied to this model can be found in \cite{dynamics} 
where the validity of the perturbative models considered here is assessed. 
It is usually used as the first example when new methods are developed 
\cite{bofin,rwc,Lambert2019,tempo,rivas_thermo,smirne}, here we use it as the 
first example and show that numerical integration and our approximation yields the 
same result. The Hamiltonian of the model considered is given by  
\begin{equation}\label{eq:noneq-spin}
     H = \underbrace{\frac{\omega_{0}}{2} \sigma_{z} + \frac{\Delta}{2} \sigma_{x}  }_{H_S} + \underbrace{\sum_{k} w_{k} a_{k}^{\dagger} a_{k}}_{H_B} + \underbrace{\sum_{k} g_k \sigma_{z} (a_{k}+a_{k}^{\dagger})}_{H_I}.
\end{equation}
The calculations in this section have been done in the weak coupling regime with
the low temperature where the Mean-Force correction to the Hamiltonian is negligible 
\cite{meanforce,meanforcereview}, so that the master equations have
small error in the steady state. The exponents in this section were 
obtained using ESPIRA. 

\begin{figure}[H]
    \centering
\begin{overpic}[width=\linewidth]{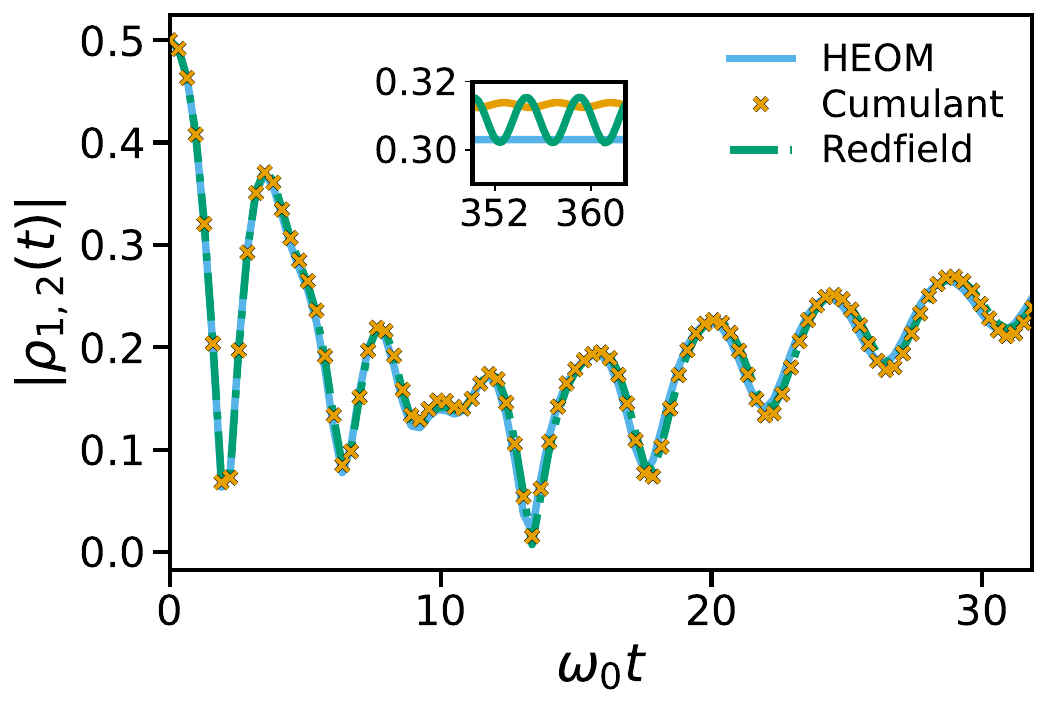}
    \put (-3,55) {\large$\displaystyle (a)$}
\end{overpic}    

\begin{overpic}[width=\linewidth]{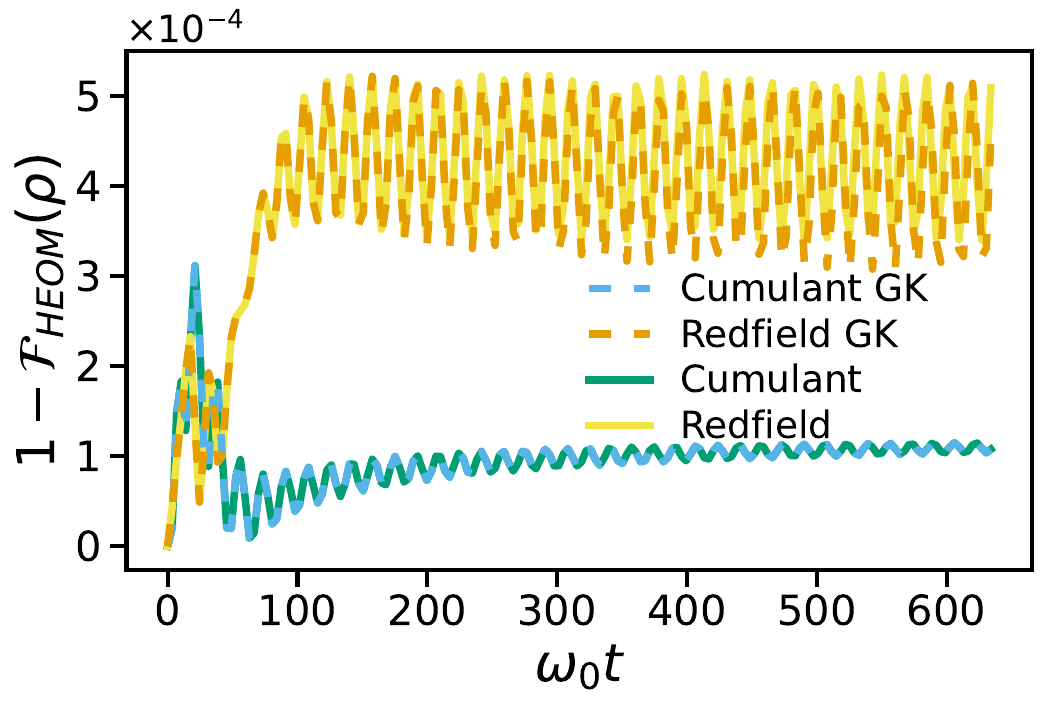}
    \put (-3,55) {\large$\displaystyle (b)$}
\end{overpic}    
 \begin{overpic}[width=\linewidth]{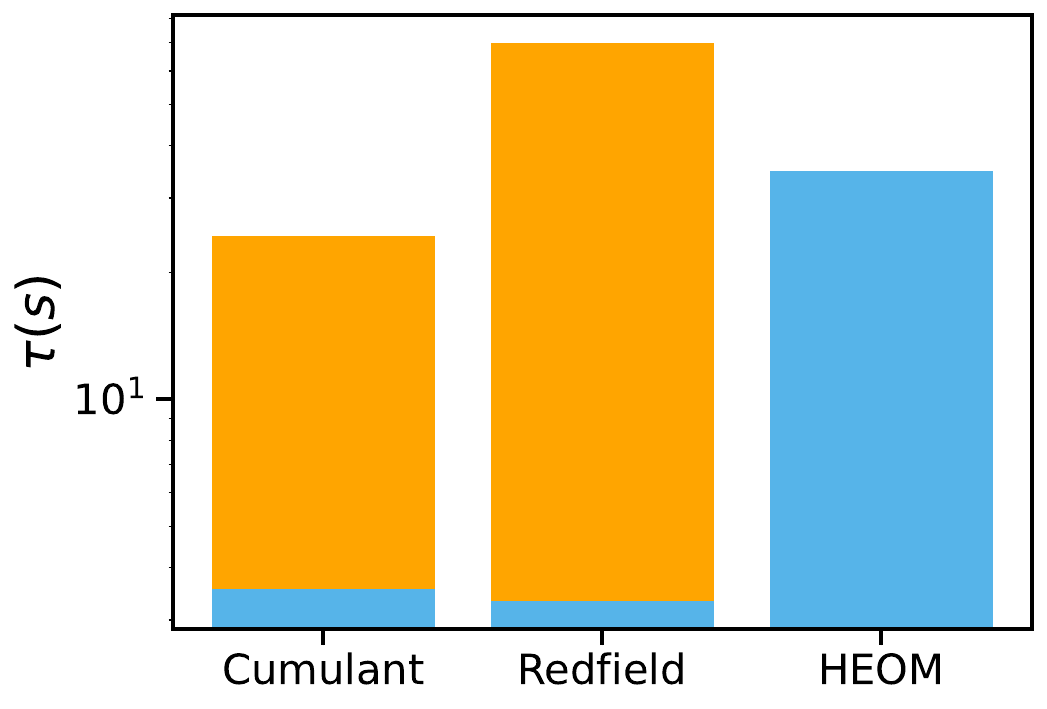}
    \put (-2,55) {\large$\displaystyle (c)$}
\end{overpic}       
\caption{ Dynamics of a spin boson model in an environment modeled with a 
Super-Ohmic spectral density with parameters 
 $T=\frac{\omega_{0}}{2}$, $s=\frac{3}{2}$, $\alpha=0.05\pi$, $\Delta=
\omega_{0}=\omega_{c}$, 
HEOM Hierarchy was set to 3, the same exponents were used for the HEOM and master 
equation simulations.
a)Absolute value of coherence.
b) Fidelity of the master equations to the numerically exact 
approach, notice the fidelity is almost one, indicating that 
the master equations are almost exact in this regime. Notice there's no difference between 
the ``GK" where the master equations were computed with Gauss-Kronrod 
integration, and the ones where we used the decaying 
exponentials approach. c) Shows the time it takes for each approach, 
the blue color indicates the usage of the decaying exponentials, while 
orange denotes the usage of the Gauss-Kronrod quadrature.}\label{fig:spin_boson}
\end{figure}

As Fig. \ref{fig:spin_boson} (b) shows the master equations are almost as precise 
as the HEOM solution, since the fidelity is close to one, in the inset of Fig. \ref{fig:spin_boson} (a)
we can see that even for long times, there are coherence oscillations in the approximate methods while there are no oscillations in the 
exact solution, this behaviour is typical in this system and is related to the so called steady state coherence \cite{dynamics,meanforce,ss1,ss2}.
Fig. \ref{fig:spin_boson} (b) also shows that using numerical integration
and our method yields the same results. However, they
take less computation time as Fig. \ref{fig:spin_boson} 
(c) shows.

Unlike other methods,where the usage of exponents increases the Hilbert space \cite{qutip5,Lambert2019,Numerically2020}, in master equations the cost of the simulation does not increase too much with the number of exponents (as $O(m)$ with $m$ being the number of exponents), The usage of decaying exponentials with these 
techniques could therefore prove immensely useful in the dynamics of 
open quantum systems in the weak coupling regime. 
\subsection{Example II: Heat Transport}\label{sect:example2}
In this section, we consider two qubits coupled to two thermal baths according 
to the Hamiltonian
\begin{align}
     H &= \underbrace{\frac{\omega_{c}}{2} \sigma^{(c)}_{z} + \frac{\Delta}{2} \sigma^{(c)}_{x}
     +\frac{\omega_{h}}{2} \sigma^{(h)}_{z} 
     + g \sigma^{(c)}_{x} \sigma^{(h)}_{x}}_{H_S} \nonumber
     \\ &+ \underbrace{\sum_{k,\alpha=h,c} \omega^{(\alpha)}_{k} a_{k}^{(\alpha)\dagger} a^{(\alpha)}_{k}}_{H_B} 
     + \underbrace{\sum_{k,\alpha=h,c} g_k \sigma^{(\alpha)}_{x} (a_{k}+a_{k}^{\dagger})}_{H_I}.
\end{align}
The inclusion of a tunneling term in the Hamiltonian of the qubits is done such that the Lamb-Shift of that qubit does not commute with the system
Hamiltonian in a fashion similar to \cite{dynamics,meanforce}.  The influence of Lamb-shift on heat currents for the Redfield equation is briefly discussed in Appendix \ref{app:current}, it is an area of research that remains unexplored, perhaps due to the fact that for the GKLS equation is an energy shift and the Lamb-shift has no important effect on heat. To compare the 
different approaches we use in this manuscript, it is necessary to introduce a 
definition for the heat current of the cumulant equation, similar to those 
defined in \cite{rivas_thermo},  we start from the internal energy of the 
system \cite{Kosloff_2013}
\begin{align}
    U=Tr[H_{S}(t)\rho_{S}(t)].
\end{align}
Taking the time derivative 
\begin{align}
    \dot{U}=Tr[\dot H_{S}(t)\rho_{S}(t)] + Tr[H_{S}(t)\dot\rho_{S}(t)],
\end{align}
we identified the contribution we can control (work or in this case its time derivative power) from the contribution we do not have control over (heat)
\begin{align}
    \dot{U}=\dot{W} + \dot Q,
\end{align}
where 
\begin{align}
   \dot{W} &= Tr[\dot{H}_{S}(t)\rho_{S}(t)], \\
   \dot Q &=Tr[H_{S}(t)\dot\rho_{S}(t)] =  J_{h}(t) + J_{c}(t) \label{eq:heat}
\end{align}
\begin{figure*}
\begin{overpic}[width=0.45\textwidth]{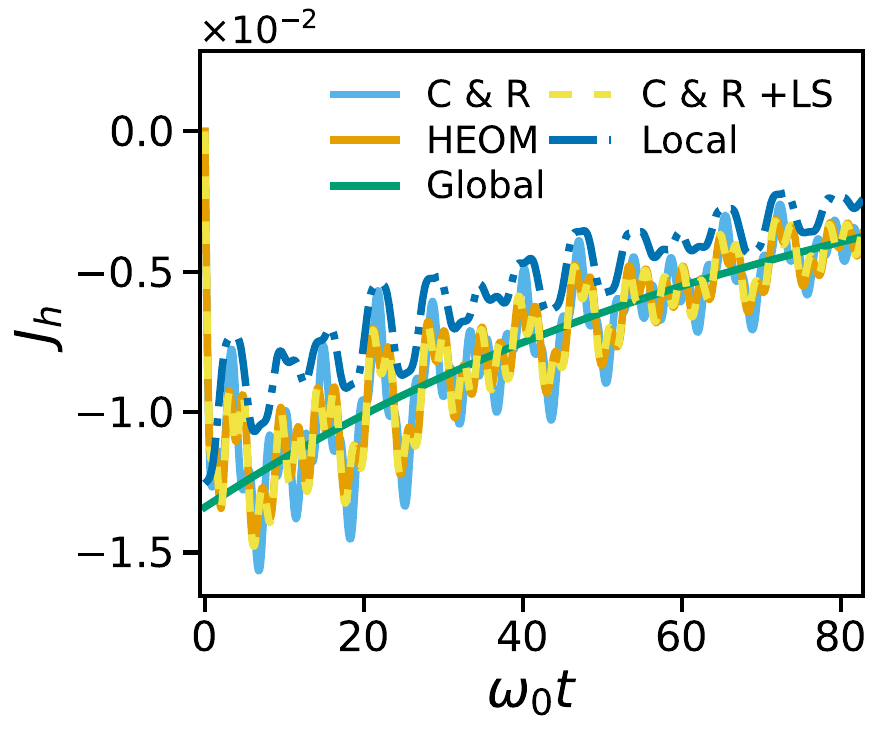}
    \put (-3,75) {\large$\displaystyle (a)$}
\end{overpic}    
\begin{overpic}[width=0.42\textwidth]{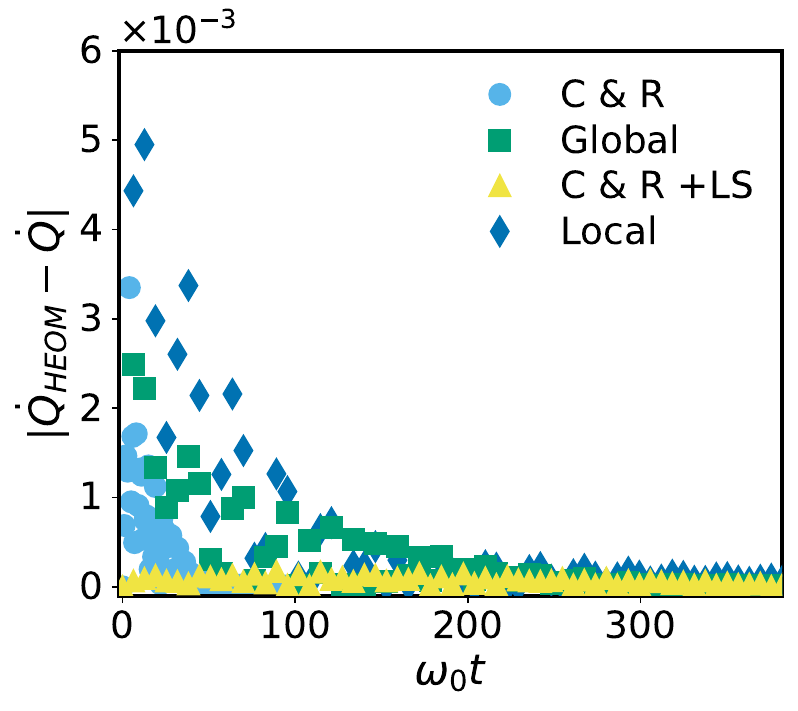}
    \put (-3,75) {\large$\displaystyle (b)$}
\end{overpic}   
\begin{overpic}[width=0.45\textwidth]{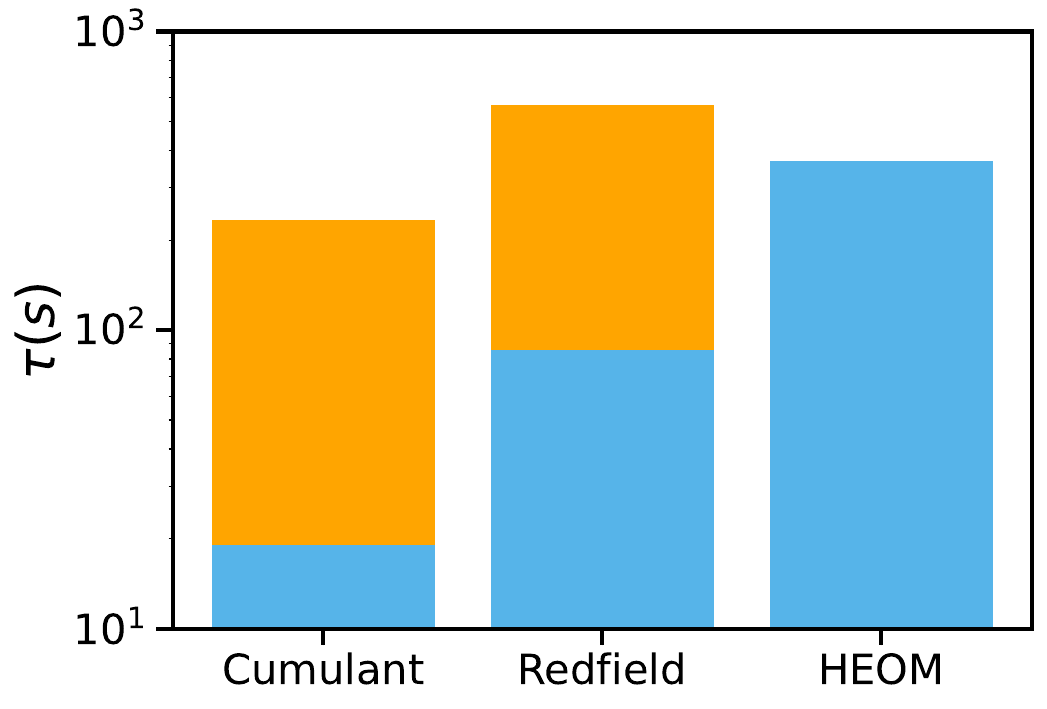}
    \put (-3,55) {\large$\displaystyle (c)$}
\end{overpic} 
\begin{overpic}[width=0.45\textwidth]{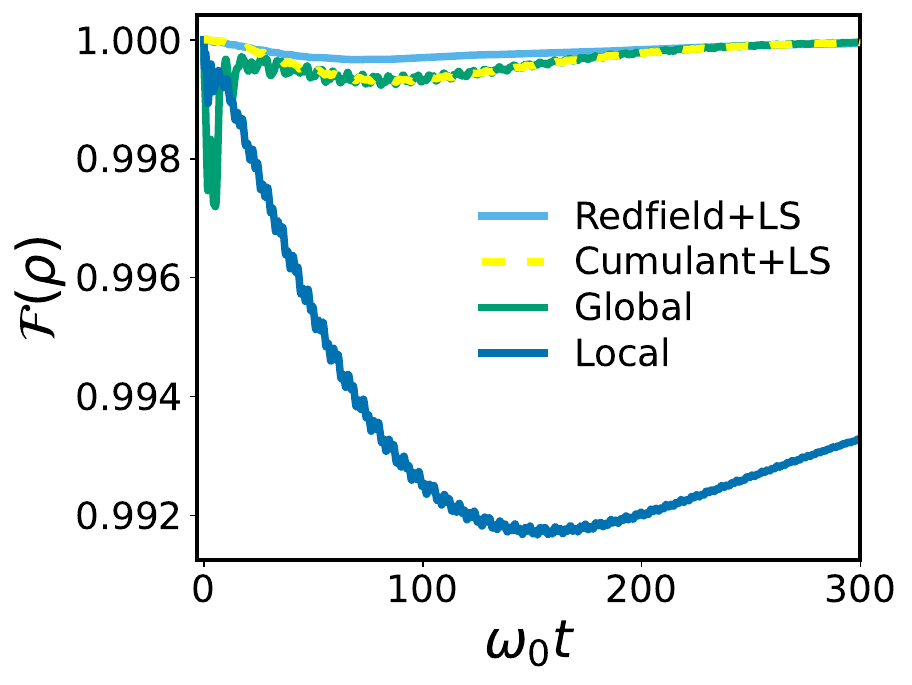}
    \put (-3,55) {\large$\displaystyle (d)$}
\end{overpic}     
\caption{\justifying Heat transport scenario, two qubits coupled to two different baths
with Underdamped spectral densities, the parameters of the simulation are $T_{c}=0.01\omega_{0}$, $T_{h}=2\omega_{0}$, $\Delta=\omega_c=\omega_h=\frac{\omega_{0}}{2}$, $g=0.5\omega_{0}$, $\Gamma=\omega_{0}$, $\lambda_c^{2}=\lambda_h^{2}=0.05\pi\omega_{0}^{3}$,
a) Shows the heat current from the environment at the hotter temperature, 
it illustrates that the Lamb-shift is necessary to obtain a proper description 
of heat currents in the transient regime and the failure of GKLS at early times
b)Shows the relative error of each method on the computation of $\dot{Q}$ with respect to HEOM. c) Shows the time used to obtain the evolution of the 
density matrix in each approach, the orange block indicates the usage 
of numerical integration, while blue indicates the usage of decaying exponentials
d) Shows the fidelity of the approaches to the 
the exact solution, notice that even though the global master equation
describes the states correctly, the description of heat currents fails as
a) and b) show.}\label{fig:heat}
\end{figure*}
where $J_h$ and $J_c$ are parts of heat currents coming from hot and cold bath, and they are obtained from relevant parts of the generator (see Appendix \ref{app:current}). 
In the case of master equations $\dot{\rho_{S}}(t)$ is easy to obtain, however in the case of the cumulant equation, which is a dynamical map, we then obtain the time derivative via
\begin{align}
    \rho_{S}(t) &= \Lambda_{t} \rho_{S}(0), \\  
    \dot{\rho_{S}}(t) &= \dot{\Lambda}_{t} \rho_{S}(0).
\end{align}
We can obtain the derivative of the map using numerical differentiation; however, 
when the coupling to the system is weak enough, one might use the fact that 
to the lowest order in coupling the time derivative of the cumulant equation 
map is the Redfield generator \cite{meanforce},
and thus the cumulant equation 
currents can be found by using the Redfield generator, the validity of this was 
checked by numerically differentiating the cumulant generator and comparing to using the Redfield 
generator, using the Redfield generator is preferable when applicable 
to avoid stability issues of numerical differentiation \cite{Quarteroni2007}.
The approximation of decay rates via decaying exponentials should enable the usage 
of automatic differentiation routines, which we hope to implement for the 
generator of non-Markovian master equations in the future. In Figure 
\ref{fig:heat}, we can see how a correct description of heat currents in time 
requires the inclusion of Lamb-shift. This begs for the investigation of the Lamb-shift 
corrections to heat currents, a previously unexplored area, warranting further 
research. Furthermore, this indicates that there could be a substantial 
influence of the Lamb shift on work which is poised to significantly impact 
the field of quantum thermal machines. Consequently, using decaying exponentials 
to compute Lamb-shift is
expected to be instrumental in advancing the understanding of heat and power 
within non-equilibrium quantum thermodynamics. On the other hand, notice from 
Figure \ref{fig:heat} shows that the GKLS master equation, both in its global and 
local forms, fails to capture the behavior of heat currents at early times, 
predicting non-zero heat at $t=0$ this is expected as they are meant as a long 
time approximation. non-Markovian master equations correctly capture this 
behavior, as other methods do \cite{popovic,menczel}.
\subsection{Example III:  A Kerr Non-linearity}\label{sect:example3}
\begin{figure}[H]
    \centering
\begin{overpic}[width=\linewidth]{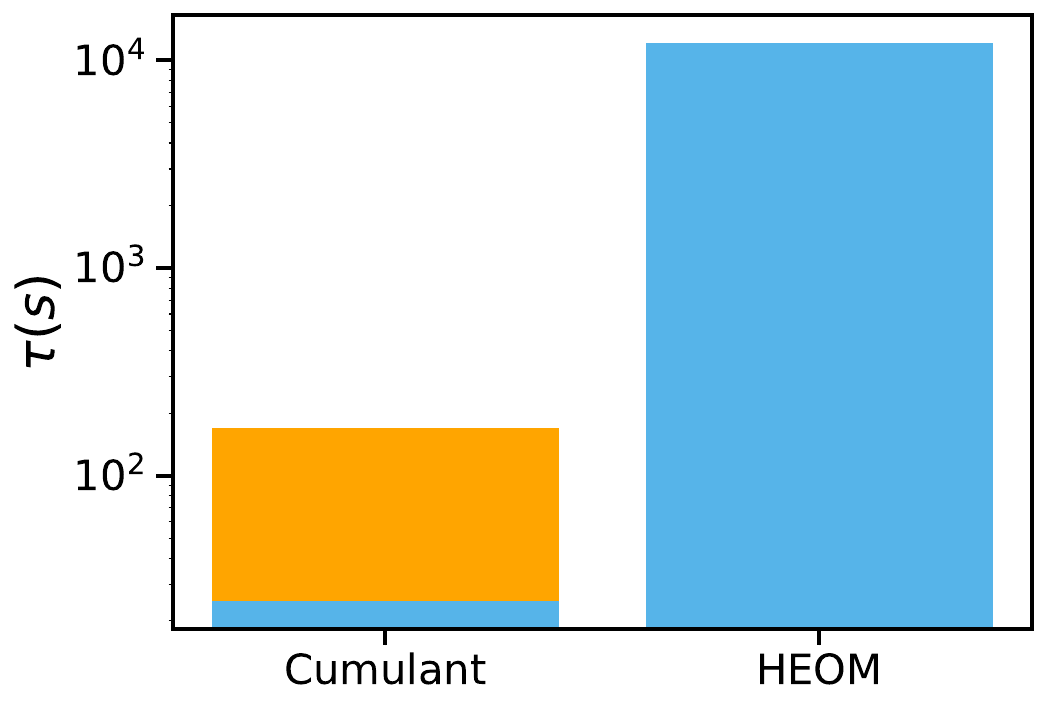}
    \put (-3,55) {\large$\displaystyle (a)$}
\end{overpic}    
\begin{overpic}[width=\linewidth]{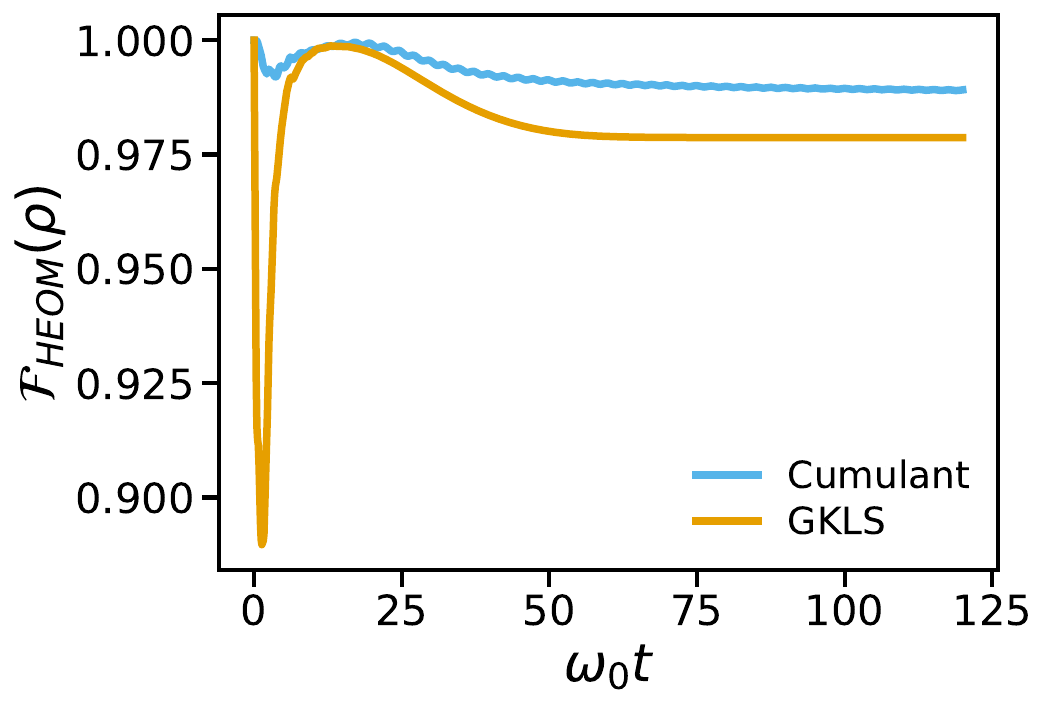}
    \put (-3,55) {\large$\displaystyle (b)$}
\end{overpic}     
\begin{overpic}[width=\linewidth]{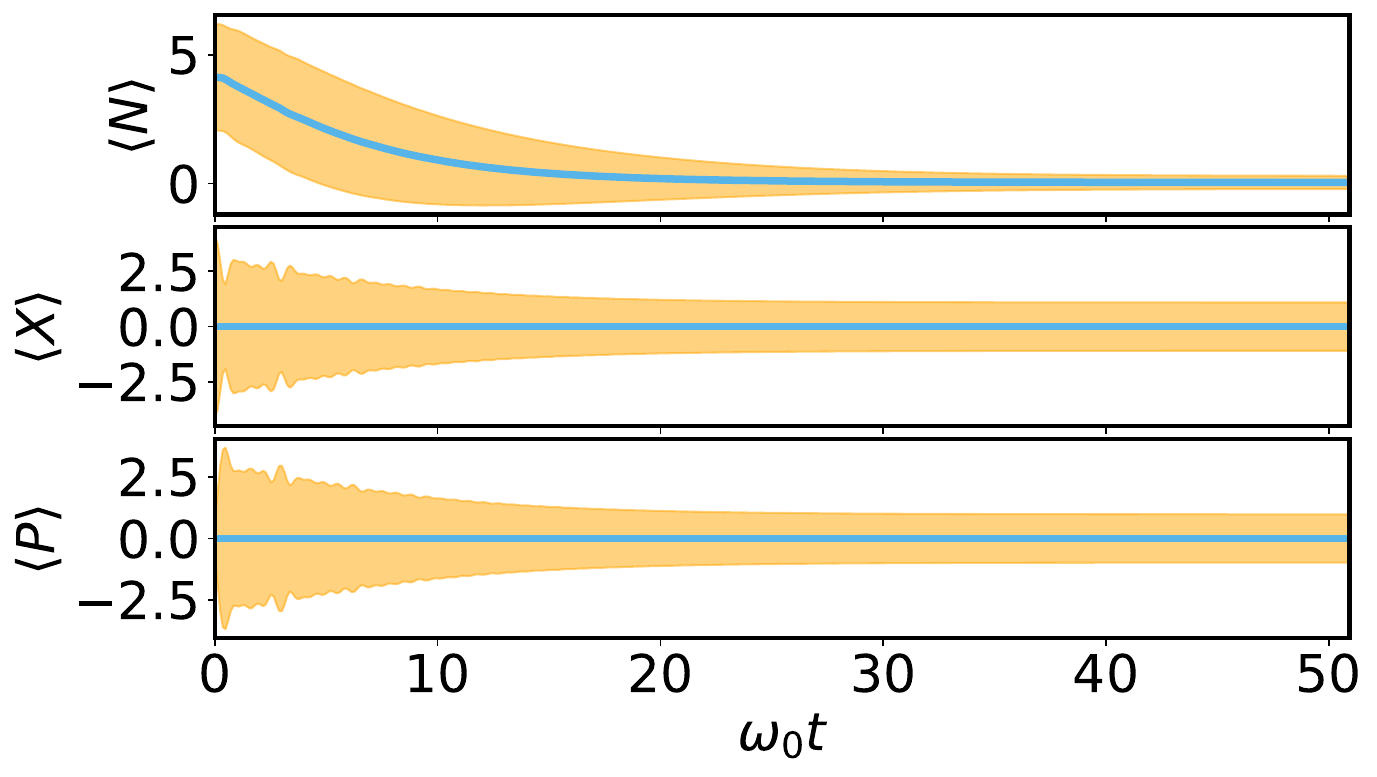}
    \put (-3,45) {\large$\displaystyle (c)$}
\end{overpic} 
\caption{Behavior of a harmonic oscillator with a Kerr Nonlinearity in a 
bath with an underdamped spectral density. a) shows the computation time required
for the cumulant and HEOM; b) shows the fidelity of the cumulant and GKLS master 
equation to HEOM; c) shows the evolution of the average population, position, and momentum
which are represented by the solid line, while the shadow represents the variance. 
The parameters used for the simulation are $T=0.1\omega_{0}$, $\lambda^{2}=\frac{\pi\omega_{0}^{3}}{4}$, $s=1$, $\omega_{c}=\omega_{0}$, $\xi=\frac{\omega_{0}}{2}$.}\label{fig:wigner}
\end{figure}

Nonlinear processes are important in quantum optics \cite{Leonhardt_2010}. 
One of the most studied processes are Kerr Nonlinearities, modeling their 
interaction with the environment is the reason the cumulant/refined weak coupling
equation was first proposed \cite{Alicki89}, the idea behind proposing this equation,
was that it could handle small spacing in Bohr frequencies, which is important 
for transient dynamics, and that it also has the GKLS master equation as 
a limit when $t \to \infty$ \cite{Alicki89,intermediate,rivas} (see 
\cite{intermediate} for an in-depth discussion ). In this section, 
we consider the Hamiltonian
\begin{align}\label{eq:kerr}
     H &=\underbrace{\omega_{0} a^{\dagger}a+\frac{\chi}{2} a^{\dagger}a^{\dagger}aa}_{H_S} \nonumber \\&+ \underbrace{\sum_{k} w_{k} a_{k}^{\dagger} a_{k}}_{H_B} + \underbrace{(a+a^{\dagger})\sum_{k} g_k (a_{k}+a_{k}^{\dagger})}_{H_I}.
\end{align}
and an underdamped spectral density. We show how the cumulant equation is almost
exact but cheaper than the numerically exact approach. To perform the simulation, 
we truncated the Hilbert space of the oscillator such that it contains $N=10$ levels. 
This number is chosen to keep HEOM feasible, but as we see from the figure, 
it is enough as the mean population of photons is significantly lower, we can also appreciate 
the difference in simulation time and precision of the different approaches. Notice from Figure \ref{fig:wigner} b) and c) that the cumulant equation has not converged to the GKLS equation,  even though the observables have seemingly reached equilibrium. This is due to the small Bohr frequencies having a long radius of convergence of the approximation \cite{Blanes_2009}. This is a phenomenon that happens often in high dimensional systems with small spacing. This simple example is proof that the cumulant equation can be an accurate description even though it supposedly tends to the Global master equation in the long time limit.
\subsection{Beyond second order descriptions and structured environments}\label{sect:example4}
The spin boson  model with the rotating wave approximation  is one of
the few systems to have an exact description when dealing with quantum reservoirs, this exact description of
course is rather limited, in the sense that the original system underwent several approximations, nevertheless,
this exact master equation provides us with a nice example on which to test the methods we have 
developed so far with structured environments. This section follows the discussion in \cite{garraway_jc,breuer} with the 
addition of structured environments and the cumulant equation. Consider
\begin{align}
	H = \omega_{0} \sigma_{+} \sigma_{-}                     
	+ \sum_{k} \omega_{k} b_{k}^{\dagger}b_{k}             
	+ \sigma_{+} \otimes \sum_{k} g_{k} b_{k} +H.C .
\end{align}
This model is simpler than the one considered in section \ref{sect:example1},
because it preserves excitations \cite{breuer,garraway_jc}, this fact allows
us to solve the system exactly when the environment is at zero temperature. To illustrate how the exponential series can capture arbitrary spectral 
densities, in this section, we will consider a realistic spectral density 
taken from \cite{lorenzoni}, the details of the spectral density are provided in Appendix \ref{app:spectral_density}, that spectral density was rescaled, so that the 
rotating wave approximation is valid.

On the other hand, the time convolutionless equation $TCL$ takes a simple form,
for this model, allowing us to illustrate the usage of the decaying exponential 
approximation in higher-order master equations, which are often not used 
due to the high-dimensional integrals in them, in a spirit similar to that in 
Crowder et al.\cite{Crowder_2024}, who reduced the dimensionality of integrals 
for Ohmic environments. We aim to overcome that limitation using our 
approximation which works for arbitrary environments. The $TCL$ equation to 
arbitrary order is given by \cite{breuer,smirnejc}

\begin{align}\label{eq:exact_me}
\dot{\rho}_{s}(t) & =-i \frac{S(t)}{2}[\sigma_{+}\sigma_{-},\rho_{s}(t)] \nonumber\\&+ \gamma(t) \left(\sigma_{-} \rho_{s}(t)\sigma_{+}-  \frac{1}{2}\{\sigma_{+}\sigma_{-}, \rho_{s}(t)\}\right),
\end{align}
where $\gamma(t)$ and $S(t)$ can be obtained from \cite{breuer,smirne}
\small
\begin{align}\label{eq:tcl_exact_me_expansion}
	\frac{\gamma(t)+ i S(t)}{2}  &=\sum_{n=1}^{\infty} (-1)^{n}\alpha^{2 n } \int_{0}^{t} dt_{1}\int_{0}^{t_{1}} dt_{2} \dots \int_{0}^{t_{2n-2}} dt_{2n-1} \nonumber
 \\&\times \left(f(t-t_{1})f(t_{1}-t_{2})\dots f(t_{2n-2}-t_{2n-1})\right)_{oc},
\end{align}
\normalsize
where $f(\tau)=C(\tau)e^{i \omega_{0} \tau}$ and oc stands for ordered cumulant.
The $TCL_{2n}$ equation corresponds to truncating the sum to $n$, $n=1$ corresponds to the 
Redfield equation we have been studying so far. while $n=2$
yields

\begin{figure*}
    \centering
\begin{overpic}[width=0.45\linewidth]{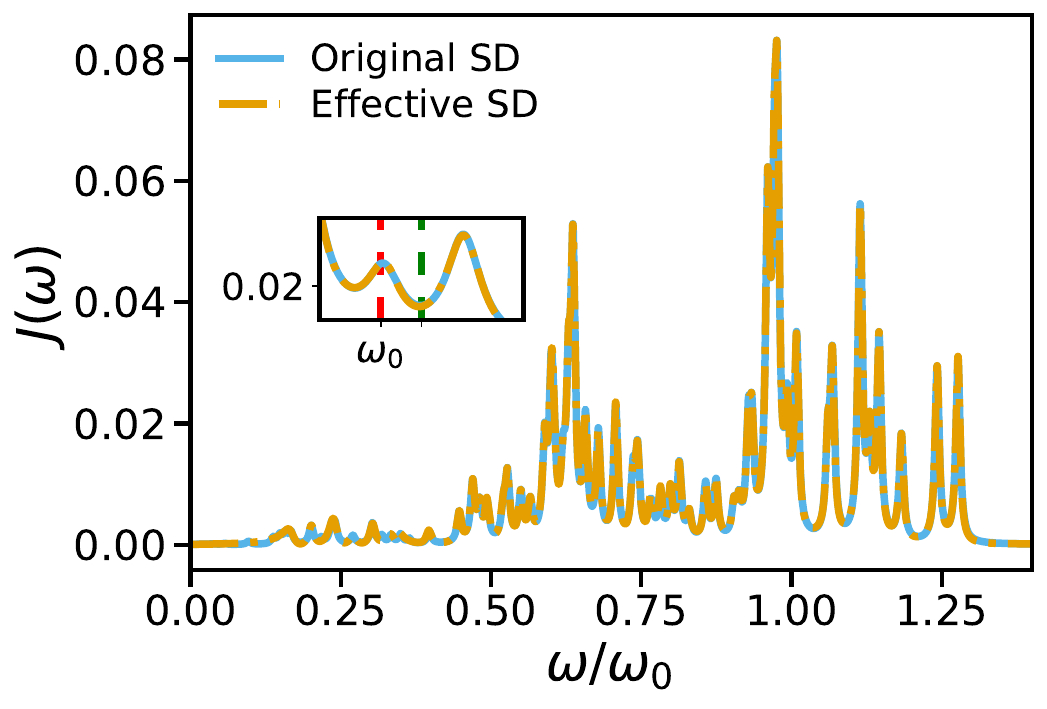}
    \put (-3,55) {\large$\displaystyle (a)$}
\end{overpic}    
\begin{overpic}[width=0.45\linewidth]{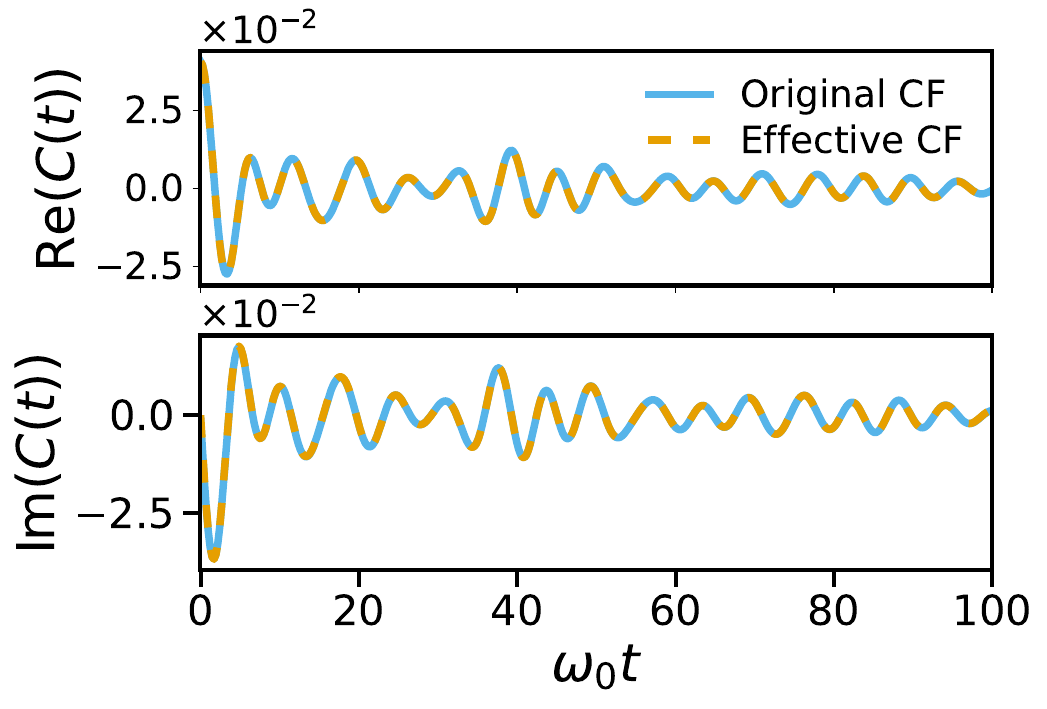}
    \put (-3,65) {\large$\displaystyle (b)$}
\end{overpic}     
\begin{overpic}[width=0.45\linewidth]{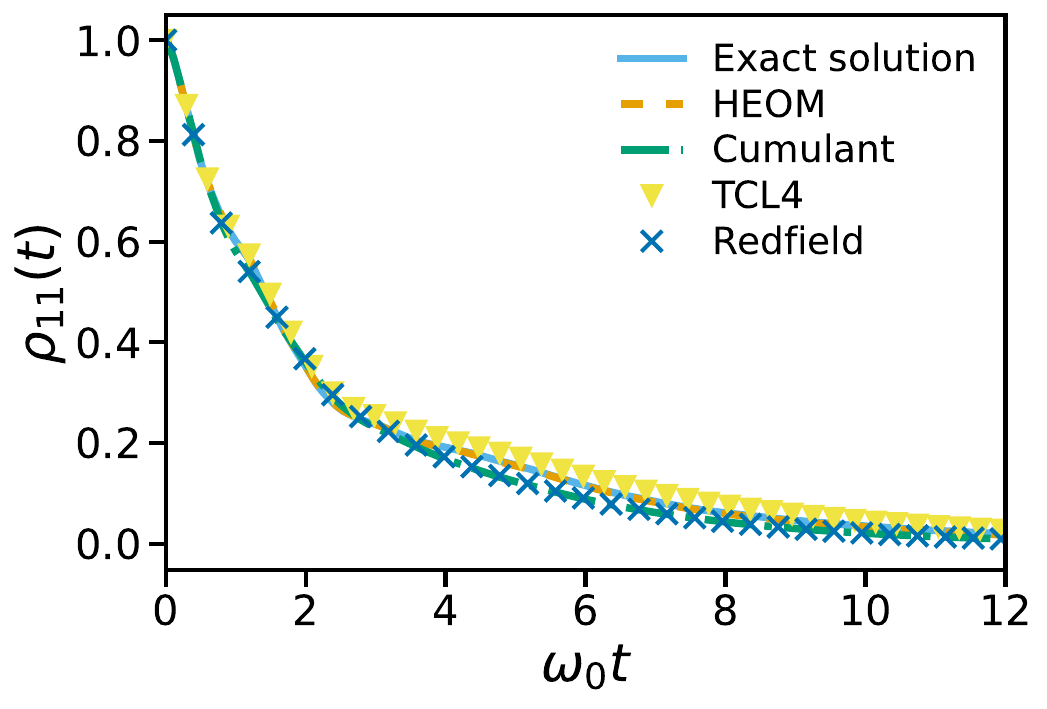}
    \put (-3,55) {\large$\displaystyle (c)$}
\end{overpic} 
\begin{overpic}[width=0.45\linewidth]{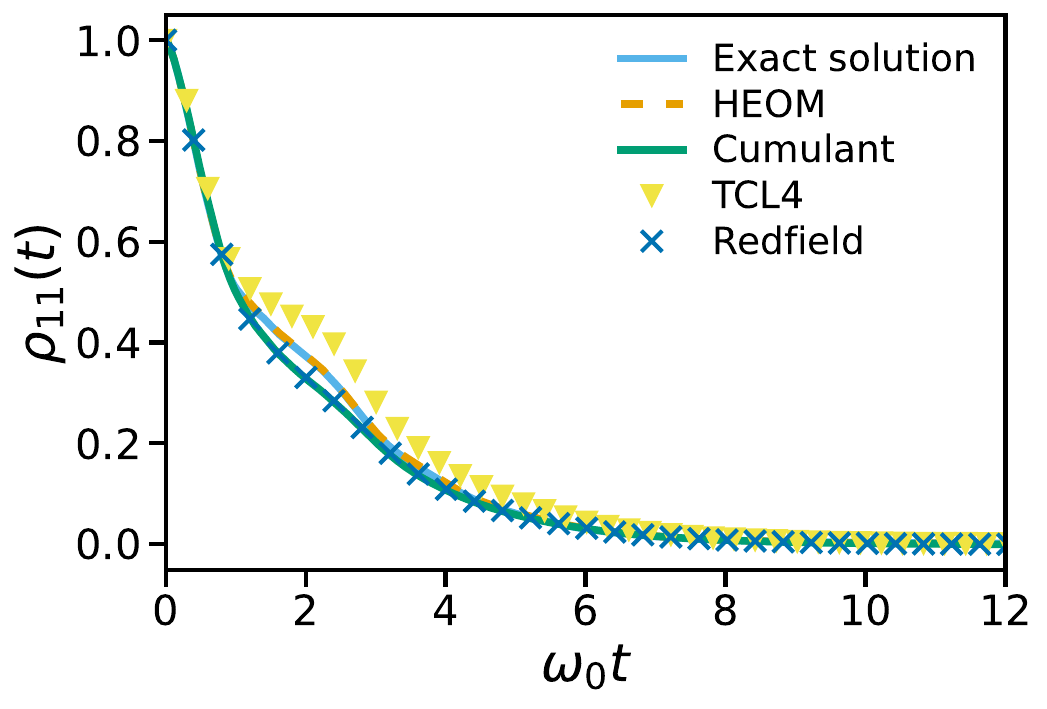}
    \put (-3,55) {\large$\displaystyle (d)$}
\end{overpic} 
\caption{\justifying a) Spectral density under consideration, the dashed line
is the effective spectral density, namely the spectral density that corresponds 
to the decaying exponential approximation, the inset shows the spectral density
near the frequency of the two-level system, b) The correlation function
under consideration, again the effective one refers to the approximation.
c) Decay of the population with all the different methods. The effective enrionment was obtained with $50$ exponents using ESPIRA}\label{fig:jc}
\end{figure*}
\small
\begin{align}\label{eq:tcl_me_4}
\left(\frac{\gamma(t)+ i S(t)}{2}\right)_{4} &= \int_{0}^{t} dt_{1}\int_{0}^{t_{1}} dt_{2} \int_{0}^{t_{2}} dt_{3} \Big(f(t-t_{2})f(t_{1}-t_{3}) \nonumber \\&+ f(t-t_{3}) f(t_{1}-t_{2}) \Big),
\end{align}
\normalsize
where $\gamma$ is the $4^{th}-order$ contribution to the decay rate and $S(t)$ to the 
Lamb-shift
\small
\begin{align}
&\left(\frac{\gamma(t)+ i S(t)}{2}\right)_{4} =  \sum_{\substack{k_1, k_2 \\ k_1 \neq k_2}} \frac{c_{k_1} c_{k_2}}{(i v_{k_1} + \omega_0)(i v_{k_2} + \omega_0)} \nonumber \\
& \quad \times \left[ \frac{e^{-t(v_{k_2} - i \omega_0)}}{v_{k_1} - v_{k_2}} + \frac{e^{-t(v_{k_1} + v_{k_2} - 2i \omega_0)}}{v_{k_2} - i \omega_0} + \frac{1}{i \omega_0 - v_{k_1}} \right. \nonumber \\
& \left. \qquad + \left(t + \frac{1}{v_{k_1} - i \omega_0} - \frac{1}{v_{k_2} - i \omega_0} + \frac{1}{v_{k_2} - v_{k_1}}\right) e^{-t v_{k_1} + 2 i t  \omega_0} \right] \nonumber \\
& + \sum_k \frac{c_k^2}{(v_k - i \omega_0)^3} \left[ 1 - e^{-2t(v_k - i \omega_0)} - 2t(v_k - i \omega_0)e^{-t(v_k - i \omega_0)} \right].
\end{align}
\normalsize
Notice that the decay rates in previous sections were approximated by a sum of 
$m$ elements, where m is the number of exponents we used to describe our two 
time correlation function, as the sum was over only one index the computational 
cost was $O(m)$. However, from the $TCL_{2n}$ expansion with ordered cumulants. 
We see that after approximation we will have $O(m^{n})$ indices in the summation (as it will be an integral of the product of $n$ correlation functions),
increasing the computational cost, thus for higher-order master equations, 
reducing the number of exponents used  has a significant impact on the 
computational cost of the simulation.

This model also allows us to illustrate how one may use the approximation of the 
two-time correlation function to solve integrodifferential Volterra equations \cite{Gao2022,VABISHCHEVICH2022177},
the exact dynamics for this model is given by \cite{breuer} 
 \begin{align}
	\dot{c_{1}}(t) =-  \int_{0}^{t}ds c_{1}(s)f(t-s),
\end{align}
such that 
\begin{align}
    \rho_{11}(t)= |c_{1}(t)|^{2}.
\end{align}
Which can be mapped to a system of ODES with ($m$+1) equations. As shown in 
Appendix \ref{app:volterra}. Thus our approximation may prove useful in solving this 
sort of equation \cite{Gao2022,VABISHCHEVICH2022177} which often appears in the context of non-equilibrium green functions
and open systems using the Mori-Zwanzig formalism 
\cite{volterra1,volterra2,volterra3,volterra4,volterra5}.

We can observe from Figure \ref{fig:jc} c) that in the weak coupling regime, the 
non-Markovian master equations resemble the exact solution, even when 
structured spectral densities are involved, the next order of the time 
the convolutionless equation does capture the increase in population better 
than the second-order equations but seems to deviate a bit more from the 
exact solution, similar to the discussion for high temperature in \cite{chentcl},
if we change $\omega_{0}$ as in d) $TCL_{4}$ follows the exact solution more 
closely, The relative performance between Redfield and fourth order 
$TCL_{4}$ master equation is determined by  the value of $J(\omega_{0})$.
This observation is consistent with the discussion in \cite{chentcl}, 
indicating that $TCL_{4}$ is less accurate under strong driving conditions. This suggests a promising research direction in exploring higher-order $TCL_{2n}$
equations for systems coupled to structured environments, which is now possible
using a series of decaying exponentials.

\section{Conclusions}

We have proved the value of using approximation of the correlation function to perform 
simulations of the dynamics of open quantum systems using non-Markovian master equations, 
we prove we obtain significant speed ups while not sacrificing any accuracy. The approximation
of the correlation function allow us to write the decay rates and Lamb-shift corrections of 
non-Markovian master equations as algebraic expressions, allowing us to compute them in  $O(m^{\frac{n}{2}})$
where $m$ is the number of exponents used to approximate the correlation function and $n$ is the 
order of the master equation used. 

We also compared the different existent methods in the literature to obtain an approximation of the 
correlation function, and concluded that ESPIRA is the best approach for master equations where 
we put more emphasis on better approximating the decay rates and satisfying detailed balance and less 
emphasis in reducing the number of exponents as in the case of numerical exact methods \cite{Numerically2020,bofin,menczel,Lambert2019}.

Futhermore we showed the impact that the Lamb-shift correction of non-Markovian master equations 
has on heat currents, demonstrating that it must be included to correctly describe the evolution 
of heat in quantum thermal machines. Paving the way to revisit studies that study optmiality
of work and heat usingthe GKLS master equation with similar but more accurate models 
\cite{finitetime1,finitetime2,finitetime3,finitetime4,finitetime5,finitetime6,finitetime7,finitetime8,finitetime9}.

Finally, we demonstrated the usefulness of our approximation when dealing with highly structured 
spectral densities, showing good agreement between the approximate and exact solutions used. This example 
also highlights how the approximation may be used to turn integro-differential equations 
into systems of ordinary differential equations. The examples in the paper also serve to 
highlight the usefulness of the cumulant equation, which so far has only been used in simple models in the literature.

\acknowledgments{ We thank Neill Lambert and Paul Menczel, whose mentoring during G.S.'s visits to RIKEN was invaluable. G.S. acknowledges the support by the Polish National Science Centre grant OPUS-21 (No: 2021/41/B/ST2/03207). M.H. acknowledges support
by the IRA Programme, project no. FENG.02.01-IP.05-
0006/23, financed by the FENG program 2021-2027, Priority FENG.02, Measure FENG.02.01., with the support
of the FNP.}

\bibliography{main}

\break
\appendix
\onecolumngrid

\section{Turning the Integro-Differential equation into a system of ODES}\label{app:volterra}

In this section, we will briefly discuss how to turn a Volterra equation into a system of ordinary differential equations by approximating the memory kernel. While the procedure is specific to Eq. \eqref{eq:volterra} a similar strategy can be employed for other integro-differential equations. In our case of interest the dynamics is given by 
\begin{align}\label{eq:volterra}
	\dot{c_{1}}(t) =-  \int_{0}^{t}ds c_{1}(s)f(t-s).
\end{align}
Now if we assume that we can represent out environment's two time correlation 
function as a sum of decaying exponentials then 
we have
\begin{align}
	f(\tau)&= \sum_{k}^{N} c_{k} e^{-\nu_{k}\tau}, \\
	\dot{c_{1}}(t) &=- \sum_{k}^{N} \int_{0}^{t}ds c_{1}(s) c_{k} e^{-\nu_{k}(t-s)},
\end{align}
if we let 
\begin{align}\label{eq:jc_yk}
	y_{k}(t) =  \int_{0}^{t}ds c_{1}(s) c_{k} e^{-\nu_{k}(t-s)},
\end{align}
then 
\begin{align}
	\dot{c_{1}}(t) =- \sum_{k}^{N} y_{k}(t).
\end{align}
To find the $y_{k}(t)$ we simply take the time derivative of \eqref{eq:jc_yk}
resulting in 
\begin{align}
	\dot{y}_{k}(t) &= c_{k} c_{1}(t) -\nu_{k} \int_{0}^{t}ds c_{1}(s) c_{k} e^{-\nu_{k}(t-s)} \\
	&= c_{k} c_{1}(t) -\nu_{k} y_{k}(t),
\end{align}
where we used
\begin{align}
	\frac{d}{dx}\left(\int_{a(x)}^{b(x)}f(x,t) dt\right)
	=f\big(x,b(x)\big) \cdot \frac{db(x) }{dx}- f\big(x,a(x)\big)
	\cdot \frac{da(x)}{dx}+\int_{a(x)}^{b(x)}\frac{\partial}{\partial_{x}}
	f(x,t) dt.
\end{align}
The system is solved subjected to the initial conditions
\begin{align}
	y_{k}(0)=0,
\end{align}
with this change the Volterra equation can be solved as $m+1$ differential
equations, where $m$ is the number of exponents used to approximate the memory kernel. This way of usign the exponents, is more akin to pseudomodes or HEOM, since we use each exponent as an auxiliary dynamical variable. 
\section{Techniques to Approximate the correlation function by a series of decaying exponentials}\label{app:exponents}

There are several techniques that can be used to approximate the correlation function as a series of decaying exponentials, in this section, we will briefly overview the techniques that appear in figure \ref{fig:correlation_methods}, the table \ref{tab:visual_comparison} contains a comparison and general recommendations as to when to use each method.

\subsection{Approximation via Non-Linear Least squares}
Nonlinear least squares fitting is a mathematical optimization technique used to model data by finding the best-fit parameters of a nonlinear function. This approach minimizes the sum of the squared differences (residuals) between the observed data points and the corresponding values predicted by the model. It is a powerful tool for analyzing experimental data and is widely applied in fields such as physics, chemistry, biology, engineering, and finance. The goal is to find the set of parameters $\theta$
that minimizes the residual sum of squares (RSS):
\begin{align}
    RSS= \sum_{i=1}^{N} (y_{i}-f(x_{i},\theta))^{2},
\end{align}
where $y_{i}$ is the data to be fitted, $x_{i}$ the independent variable in out case time, $f(x_{i},\theta)$ the function we wish to use to explain the data, and $N$ is the number of data points. The function we fit our correlation function to is 
\begin{align}
    C(t) &=C_{R}(t)+i C_{I}(t),
\end{align}
where we've separated the real and imaginary part, and perform a different fit for each as outlined in \cite{bofin}. The real and imaginary parts are given by
\begin{align}
    C_{R}(t,c^{R},\nu^{R},\nu^{I}) = \sum_{k}^{m} c_{k}^{R} e^{-\nu_{k}^{R}} \cos(\nu_{k}^{I}), \\
    C_{I}(t,c^{I},\nu^{R},\nu^{I}) = \sum_{k}^{m} c_{k}^{I} e^{-\nu_{k}^{R}} \sin(\nu_{k}^{I}),
\end{align}
where the superscript $R,I$ refers to real and imaginary parts respectively and no subscript refers to a vector of the $m$ parameters. The main advantange of this approach with respect to the others is that it allows to constraint $c_{k}$ and $v_{k}$,  
which can often help make numerically exact solutions more stable, we refer to this way of obtaining the exponents as NLSQ-CF.

NLSQ is also flexible, we can also fit our spectral density to be a combination of underdamped spectral densities for which we know the Matsubara decomposition instead of fitting the correlation function directly \cite{meier,eheom}, in this sense the model we fit our spectral density to is \cite{qutip5,bofin}
\begin{equation}
J_{\mathrm approx}(\omega; a, b, c) = \sum_{i=1}^{m} \frac{2 a_i b_i \omega}{((\omega + c_i)^2 + b_i^2) ((\omega - c_i)^2 + b_i^2)}.
\end{equation}
We then generate the set of exponents for each of the $m$ spectral densities using the Matsubara expansion, 
where we would have to set the number of exponents for the expansion. Inspired by this method and the AAA method, we propose fitting the power  spectrum instead (NLSQ-PS) using the model function 
\begin{align}
    S(\omega) = \sum_{k=1}^{N}\frac{2(a_k c_k + b_k (d_k - \omega))}
{(\omega - d_k)^2 + c_k^2}.
\end{align}
This allows for using NLSQ without relying on the Matsubara expansion unlike 
previous methods based on the spectral density \cite{bofin,meier,eheom}, this is a massive improvement on the spectral density fitting method \cite{eheom,meier} as this is not limited by Matsubara exponents in the zero temperature limit.  One recovers the correlation function
by realizing that the inverse Fourier transform of each individual $k$ component is given 
by
\begin{align}
    C_{k}(t) = (a_{k} + i b_{k}) e^{-(c_{k}+i d_{k})} \qquad \text{for }t>0.
\end{align}
In figure \ref{fig:correlation_methods} we can see NLSQ-PS is one of the best performing ones. The paper uses the implementation in scipy \cite{2020SciPy}, see their documentation for more details.
\subsection{The AAA method}

The Adaptive Antoulas–Anderson algorithm (AAA) is a method for the approximation of a function in terms of a rational function \cite{taming,takahashiHighAccuracyExponential2024,aaa}
\begin{align}
    f(z)  \approx \frac{q(z)}{p(z)} = \frac{\sum_{j=1}^{m} \frac{\text{weight}_j f_j}{z - z_j}}{\sum_{j=1}^{m} \frac{\text{weight}_j}{z - z_j}} =\sum_{j=1}^{m} \frac{residues}{z-poles},
\end{align}
where the $z_{j}$ are called support points and the $f_{j}$ are the values of the function at support points.
These sort of expressions are called rational barycentric approximations \cite{aaa}. We don't use this method on the correlation function directly, but on the power spectrum which we obtain via the fluctuation-dissipation theorem \cite{taming,fermionicaaa}.
After obtaining this rational polynomial form of the power spectrum one can recover the correlation function by noticing that 
\begin{align}\label{eq:AAA_power}
    S(\omega) = \int_{-\infty}^{\infty} dt e^{i \omega t} C(t)  = 2 \Re \left(\sum_{k} \frac{c_{k}}{\nu_{k}-i \omega} \right),
\end{align}
Which allows us to identify 
\begin{align}\label{eq:poles}
    \nu_{k}= i \times poles, \\
    c_{k} = -i \times residues.
\end{align}
For a  detailed discussion of how this method works, see \cite{aaa}, while for its application on open quantum systems see \cite{takahashiHighAccuracyExponential2024,taming,fermionicaaa}.
Here we provide a brief overview of the main steps 
\begin{enumerate}
    \item Supply  the sample set $Z= (Z_{1},\dots,Z_{N})$.
    \item Start the Loop and repeat until $|F(Z)-r(z)|<tol$ where $F$ is the original 
    function and $r$ its rational approximation.
    \item Initialize the rational approximation $r$, support points $z$, and a vector $f$ to be vectors of zeros.
    \item Find  the $k$ for which $|F(Z_{k})-r(Z_{k})|$ is maximum and add $Z_{k}$ to the support points $z$,
    add the value of the function at the support point to the vector $f$.
    \item Compute the Cauchy and matrices for $Z$ and $z$ (substraction via outer product)
    \begin{align}
        C= \frac{1}{Z-z}.
    \end{align}
    \item Compute the Loewner Matrix 
    \begin{align}
        L= \frac{F-f}{Z-z}.
    \end{align}
    \item Perform SVD on the Loewner matrix  $L=U D W$ and keep the matrix w which denotes the weights of the approximation.
    \item We get the rational approximation of the function $    r(z) = \frac{q(z)}{r(z)}$ by computing the numerator and denominator as 
    \begin{align}
    q(z)=  C (w F), \qquad     p(z) = C w .
    \end{align}
    \item We check if our rational approximation satisfies our required tolerance, it it does 
    we break the Loop $|f(Z)-r(Z)|< tol$.
    \item The Loop ran $2 N + 1$ times before breaking, compute the $2 N$, poles and residues.
    residues are obtained via the simple quotient rule on $r$
    \begin{align}
        residues =  - \frac{C (w f)}{ C^{2} w},
    \end{align}
    while poles correspond to the eigenvalues of the generalized eigenvalue problem
    \begin{align}
    \begin{pmatrix}
    0 & w_1 & w_2 & \cdots & w_m \\
    1 & z_1 &   &        &   \\
    1 &   & z_2 &        &   \\
    \vdots &   &   & \ddots &   \\
    1 &   &   &        & z_m
    \end{pmatrix}
    = \lambda
    \begin{pmatrix}
    0 &   &   &        &   \\
    & 1 &   &        &   \\
    &   & 1 &        &   \\
    &   &   & \ddots &   \\
    &   &   &        & 1
    \end{pmatrix}.   
    \end{align}    
    \item From Eq. \ref{eq:AAA_power} we see that $c_{k}$ and $\nu_{k}- i \omega$ come in complex conjugate pairs.
    to find them. We filter the poles in the lower half on the complex plane, and identify the $c_{k}$ and $\nu_{k}$
    with  Eq. \ref{eq:poles}.
\end{enumerate}

\subsection{Prony Polynomial based methods}

The Prony polynomial forms the mathematical foundation for many spectral analysis techniques that estimate frequencies, damping factors, and amplitudes of signals. These methods work by interpreting a given signal as a sum of complex exponents and deriving a polynomial whose roots correspond to the frequencies or poles of the system.

The methods consider $M$ samples a linearly spaced signal
\begin{align}
	(f(t_{0}),f(t_{0}+\delta t), \dots f(t_{0}+ (M-1) \delta t ),
\end{align}
and assume that this signal comes from a sum of complex exponents
\begin{align}
	f(t)=\sum_{k=0}^{N-1} c_{k} e^{-\nu_{k} t} =\sum_{k=0}^{N-1} c_{k} z_{k}^{t} .
\end{align}
The Prony polynomial based methods estimate the phases $z_{k}^{t}$ from which we can obtain the
$\nu_{k}$ and the amplitudes $c_{k}$. The $z_{k}$ can be seen as the generalized
eigenvalues of the matrix pencil \footnote{Typically there is another dimension
	involved in the matrix pencil an upper bound on the number of exponents, and
	the number of exponents is treated as an unknown,
	we omitted this for simplicity}\cite{esprit}
\begin{align}\label{eq:matrix_pencil}
	z_{j}  {\mathbf H}_{M-N,N}(0) - {\mathbf H}_{M-N,N}(1) = {\mathbf V}_{M-N,N}
	({\mathbf z})   \mathrm{diag}  \Big( \left( (z_{j} - z_{k})\gamma_{k}
	\right)_{k=1}^{N} \Big) {\mathbf V}_{N,N}({\mathbf z})^{T},
\end{align}
where $\mathbf H$ denotes a Hankel matrix built from the signal, the subscripts
indicating the sampling points in the first column and last row respectively,
the argument of the Hankel matrix denotes how many shifts to apply to the
signal. For example
\begin{align}
	{\mathbf H}_{3,2}(0) = \begin{pmatrix}
		                       f(t_{0})            & f(t_{0} +\delta t)   & f(t_{0} + 2\delta t) \\
		                       f(t_{0} + \delta t) & f(t_{0} + 2\delta t) & f(t_{0} + 3\delta t)
	                       \end{pmatrix}.
\end{align}

The amplitudes ($c_{k}$) can later be obtained by solving the least-squares Vandermonde system given by
\begin{align}
	V_{N,M}(z)c = f ,
\end{align}
where $V_{N,M}(z)$ is the Vandermonde matrix given by
\begin{align}
	V_{M,N}(z)=\begin{pmatrix}
		           1         & 1         & \dots  & 1         \\
		           z_{1}     & z_{2}     & \dots  & z_{N}     \\
		           z_{1}^{2} & z_{2}^{2} & \dots  & z_{N}^{2} \\
		           \vdots    & \vdots    & \ddots & \vdots    \\
		           z_{1}^{M} & z_{2}^{M} & \dots  & z_{N}^{M} \\
	           \end{pmatrix}  ,
\end{align}
$M$ is the length of the signal, $N$ the number of exponents, and $f=f(t_{sample})$ is the signal evaluated in the sampling points, while  $c = (c_{1}, \dots, c_{N})$ is a vector containing the amplitudes.
The main difference between the Prony and the Estimation
of Signal Parameters via Rotational Invariance Techniques (ESPRIT) methods is the way one obtains the roots of the polynomial, typically whether this system is solved or a low rank approximation is found for the polynomial \cite{esprit}.

For the Prony method we obtain those eigenvalues directly (typically by using
the least squares solution of the left hand side of \eqref{eq:matrix_pencil} ), while for 
ESPRIT we perform the singular value decomposition on $H(0)$ then
\begin{align}
	{\mathbf H}_{M-N,N}(0)  = U_{M-N,M-N} D_{M-N ,N} W_{N,N},
\end{align}
where U is a unitary matrix, D is diagonal and W is a square matrix.
The phases $z_{k}$ can be obtained from the eigenvalues of the
matrix \cite{esprit,takahashiHighAccuracyExponential2024}
\begin{align}
	A = (W(0)^{T})^{+} W(1),
\end{align}
where $^{+}$ denotes the Penrose pseudoinverse. The $c_{k}$ are obtained by
least squares on the Vandermonde system

\subsection{Estimation of Signal Parameters by Iterative Rational Approximation (ESPIRA)}

ESPIRA exploits the connection between parameter estimation  (Prony like methods
estimate amplitude and phase) and the rational approximation of functions. In our 
context it exploits the fact that the correlation function and power spectrum are 
fourier transforms of each other
\begin{align}
    C(\tau)=\sum_{k}c_{k}e^{\nu_{k} t } \longleftrightarrow S(\omega) = 2 \Re \left(\sum_{k} \frac{c_{k}}{\nu_{k}-i \omega} \right),
\end{align}
Intuitively, thus not strictly true, one can think of it as ESPRIT on the correlation function and AAA on 
the power spectrum merged together. As such it provides the best of both worlds providing
good fits on both quantities as Figure \ref{fig:correlation_methods} shows. For further information 
about how the methods works see \cite{esprit,Derevianko2023}. We recommend it when 
using master equations as it is the method we find most flexible, and it is relatively straightforward
to obtain a desired accuracy, whereas other methods might not improve with more exponents. The method
steps are  (taken from Algorithm 3 \cite{esprit}):
\begin{enumerate}
    \item Supply  $\mathbf{f} = (f_k)_{k=0}^{2N-1}$ equidistant samples of the signal, and the required tolerance for the rational approximation.
    \item Compute the DFT-vector $\hat{\mathbf{f}} = (\hat{f}_k)_{k=0}^{2N-1}$ with $\hat{f}_k = \sum_{j=0}^{2N-1} f_j \omega_{2N}^{kj}$ of $\mathbf{f}$ (with $\omega_{2N} := e^{-2\pi i / 2N}$).
     \item  Use AAA to compute a rational function $r_M(z)$ of (smallest possible) type $(M-1, M)$ ($M \le N$), such that
    $$ |r_M(\omega_{2N}^{-k}) - \omega_{2N}^{-k} \hat{f}_k| < tol, \quad k = 0, \dots, 2N-1. $$.
    \item Compute the rational function representation of  $r_M(z)$,
    $$ r_M(z) = \sum_{k=1}^{M} \frac{a_k}{z - z_k}, $$
    \item obtain the amplitudes by $c_k := \frac{a_k}{1-z_k^2}$, $k=1, \dots, M$.
\end{enumerate}
The best thing about this method when compared to AAA alone, is that it is a lot 
less sensitive to the support points used, the original AAA scheme is a lot more 
sensitive, though it is usually not a problem when using logarithmic instead of 
linearly separated support points \cite{fermionicaaa,taming}.
\begin{table*} 
\centering
\caption{Comparison of methods to obtain exponents}
\label{tab:visual_comparison}
\setlength{\tabcolsep}{0.2pt} 
\renewcommand{\arraystretch}{1.2} 

\begin{tabular}{
    p{1.8cm} 
    >{\centering\arraybackslash}p{1.4cm} 
    >{\centering\arraybackslash}p{1.4cm}
    >{\centering\arraybackslash}p{1.4cm} 
    >{\centering\arraybackslash}p{1.4cm} 
    >{\centering\arraybackslash}p{1.4cm} 
    >{\centering\arraybackslash}p{1.4cm}
    >{\centering\arraybackslash}p{1.4cm} 
    p{4.5cm} 
}
\toprule 
\rowcolor{HeaderBlue}
\color{white}\textbf{Method}
& \header{Works on Arbitrary Functions}
& \header{Allows for constraints}
& \header{Does not Require Extra input}
& \header{Does not require optimization}
& \header{Not Sensitive to sampling points}
& \header{Works at arbitrary Temperatures}
& \header{Works on Noisy data}
& \color{white} \textbf{Recommended when...} \\
\midrule 
\rowcolors{2}{white}{RowGray}
\textbf{NLSQ} & \cmark & \cmark & \xmark & \xmark & \xmark & \cmark & \cmark &
You have an idea about which exponents should be included \cite{Lambert2019}. \\
\textbf{AAA} & \cmark & \xmark & \cmark & \cmark & \xmark & \cmark & \cmark &
The steady-state is most important and spectral density is not too structured.\\
\textbf{Prony} & \cmark & \xmark & \cmark & \cmark & \xmark & \cmark & \xmark &
The correlation function is noiseless and long lived. \\
\textbf{Matsubara} & \xmark & \xmark & \cmark & \cmark & \cmark & \xmark & \xmark &
Doing high temperature simulations using the specific spectral densities it is available for. \\
\textbf{ESPIRA} & \cmark & \xmark & \cmark & \cmark & \xmark & \cmark & \cmark &
A general-purpose method is required. It is the main method we recommend.\\
\textbf{ESPRIT} & \cmark & \xmark & \cmark & \cmark & \xmark & \cmark & \cmark &
The correlation function is long lived. \\
\bottomrule 
\end{tabular}
\end{table*}
\section{The Cumulant Equation}\label{app:cumulant}

In this section, we derive the approximation for the decay rates and Lamb-shift of the cumulant equation presented in the text. The major difference to the other approaches in this manuscript, and the reason this derivation is slightly longer, is that the cumulant equation integral's are on a square so it contains two regions, one where $t_{1}>t_{2}$ and another where $t_{2}>t_{1}$. For that reason we cannot integrate directly as in the $TCL$ case, but need to rewrite the integrals first.
\subsection{Decay Rates in the Cumulant equation}
The decay rates in the Refined Weak Coupling equation are given by \cite{meanforce,rwc}

 \begin{equation}\label{eq:decay_time}
\Gamma(\omega,\omega',t)=\int_{0}^{t} dt_1 \int_{0}^{t} dt_2 e^{i (\omega t_{1} - \omega' t_{2})} C(t_{1}-t_{2}).
 \end{equation} 
 
Taking into account that the correlation function satisfies

\begin{equation}\label{eq:anti_simmetry}
    C(-t) =  \overline{C(t)}.
\end{equation}

It is convenient to split the integral into the regions with $t_{1}>t_{2}$,$t_{2}>t_{1}$, as follows

 \begin{align}\label{eq:gamma_subs}
\Gamma(\omega,\omega',t) &=\int_{0}^{t} dt_1 \int_{0}^{t_{1}} dt_2 e^{i (\omega t_{1} - \omega' t_{2})} C(t_{1}-t_{2}) + \int_{0}^{t} dt_2 \int_{0}^{t_{2}} dt_1 e^{i (\omega t_{1} - \omega' t_{2})} C(t_{1}-t_{2}),
 \end{align} 
then we make the change of labels in the second integral to incorporate \eqref{eq:anti_simmetry}
\begin{align}\label{eq:decay_decompose}
\Gamma(\omega,\omega',t)&=\int_{0}^{t} dt_1 \int_{0}^{t_{1}} dt_2 e^{i (\omega t_{1} - \omega' t_{2})} C(t_{1}-t_{2}) + \int_{0}^{t} dt_1 \int_{0}^{t_{1}} dt_2 e^{i (\omega t_{2} - \omega' t_{1})} C(t_{2}-t_{1}) \nonumber \\
   &=\int_{0}^{t} dt_1 \int_{0}^{t_{1}} dt_2 e^{i (\omega t_{1} - \omega' t_{2})} C(t_{1}-t_{2}) + \int_{0}^{t} dt_1 \int_{0}^{t_{1}} dt_2 e^{i (\omega t_{2} - \omega' t_{1})} \overline{C(t_{1}-t_{2})}.
 \end{align} 
By substituting the exponential representation of the correlation function \eqref{eq:exps} and separating the terms in the sum one might write
 \begin{align}
 \Gamma(\omega,\omega',t)  &=\sum_{k} \Gamma_{k}(\omega,\omega',t) = \sum_{k}c_{k}\int_{0}^{t} dt_1 \int_{0}^{t_{1}} dt_2 e^{i (\omega t_{1} - \omega' t_{2})} e^{- \nu_{k} (t_{1}-t_{2}) } \nonumber
 + \sum_{k}\overline{c_{k}}\int_{0}^{t} dt_1 \int_{0}^{t_{1}} dt_2 e^{i (\omega t_{2} - \omega' t_{1})} e^{- \overline{\nu_{k}} (t_{1}-t_{2}) },
 \end{align} 
thanks to the exponential representation of the correlation function the integrals are straight-forward, carrying up the integration yields
\begin{align}\label{eq:cum_diff}
\Gamma_{k}(\omega,\omega',t)   &= \frac{c_{k} e^{-(\nu_{k}-i\omega)t}}{(\nu_{k}-i \omega)(\nu_{k}-i \omega')}+\frac{\overline{c_{k}} e^{-(\overline{\nu_{k}}+i\omega')t}}{(\overline{\nu_{k}}+i \omega)(\overline{\nu_{k}}+i \omega')} \nonumber \\
&+ \frac{i c_{k}}{\omega-\omega'} \left( \frac{1}{(\nu_{k}-i \omega)} - \frac{e^{i(\omega-\omega') t}}{(\nu_{k}-i \omega')} \right)
+ \frac{i\overline{c_{k}}}{\omega-\omega'} \left( \frac{1}{(\overline{\nu_{k}}+i \omega')} - \frac{e^{i(\omega-\omega') t}}{(\overline{\nu_{k}}+i \omega)} \right) 
\end{align}
Let us define 
\begin{align}
    \chi_{k}(\omega,\omega',t) &= \frac{c_{k} e^{-(\nu_{k}-i\omega)t}}{(\nu_{k}-i \omega)(\nu_{k}-i \omega')} + \frac{i c_{k}}{\omega-\omega'} \left( \frac{1}{(\nu_{k}-i \omega)} - \frac{e^{i(\omega-\omega') t}}{(\nu_{k}-i \omega')} \right), 
\end{align}
then  we can write the decay rates as 
\begin{align}
\Gamma_{k}(\omega,\omega,t) &=  \overline{\chi_{k}}(\omega',\omega,t) + \chi_{k}(\omega,\omega',t)
\end{align}
we then take the limit $\omega \to \omega'$
\begin{align}\label{eq:cum_same}
\Gamma_{k}(\omega,\omega,t) &=\lim_{\omega \to \omega'}\Gamma_{k}(\omega,\omega',t) =2 \Re\left(\chi_{k}(\omega,\omega,t)\right) =2 \Re\left(\frac{c_{k}\left((\nu_{k} t - i \omega t- 1 )+e^{-(\nu_{k}-i\omega) t}\right)}{(\nu
_{k}-i\omega)^{2}} \right).
\end{align}
These expressions substitute the standard approach which consists in taking the Fourier transform of Eq. \eqref{eq:decay_time} \footnote{The principal value is a consequence of \eqref{eq:anti_simmetry}, as imposing this condition will later result on the Fourier transform of the sign function} to obtain
\begin{align} \label{eq:Decays}
&\Gamma(\omega,\omega',t) = \frac{t^{2}}{\pi}\int_{0}^{\infty} d\nu e^{i\frac{\omega-\omega'}{2} t} J(\nu) \left[ (n(\nu)+1) \sinc\left(\frac{(\omega-\nu)t}{2}\right) \sinc\left(\frac{(\omega'-\nu)t}{2}\right)   + n(\nu) \sinc\left(\frac{(\omega+\nu)t}{2}\right) \sinc\left(\frac{(\omega'+\nu)t}{2}\right)   \right],
\end{align}
where we assumed that the bath is bosonic and in thermal equilibrium. $n(\nu)$ denotes the Bose-Einstein distribution evaluated at frequency $\nu$. Notice that even when the integral is one-dimensional, it needs to be computed for every value of $t$, thus having an algebraic expression such as Eqs. \eqref{eq:cum_diff} and \eqref{eq:cum_same} beneficial, especially when high accuracy in integration is needed. Figure \ref{fig:cum_time} shows how it is advantageous to obtain decay rates using exponential decompositions of the environment.

\subsection{ Lamb-shift in the Cumulant equation}
For the cumulant equation the Lamb-shift is given by
\begin{eqnarray}
\Lambda(t)= \sum_{w,w'} \sum \xi(\omega,\omega',t)A^{\dagger}(\omega) A(\omega'),
\end{eqnarray}
where
\begin{align}
\xi(\omega,\omega',t)=  \frac{1}{2i} \int_{0}^{t} dt_{1}\int_{0}^{t} dt_{2} \text{sgn}(t_1-t_2) e^{i(\omega t_1- \omega' t_2)}   C(t_{1}-t_{2}).
\end{align}
The sign function can be written as $\text{sgn}(x)=2\theta(x)-1$ such that 
\begin{align}
\xi(\omega,\omega',t)&=  \int_{0}^{t} dt_{1}\int_{0}^{t} dt_{2}  \frac{(2 \theta(t_{1}-t_{2})-1)}{2i} e^{i(\omega t_1- \omega' t_2)}   C(t_{1}-t_{2}), \\
&=\frac{1}{i} \int_{0}^{t} dt_{1}\int_{0}^{t} dt_{2}  \theta(t_{1}-t_{2}) e^{i(\omega t_1- \omega' t_2)}   C(t_{1}-t_{2}) - \frac{1}{2i} \int_{0}^{t} dt_{1}\int_{0}^{t} dt_{2}  e^{i(\omega t_1- \omega' t_2)}   C(t_{1}-t_{2}).
\end{align}
But notice that for the cumulant equation 
 \begin{equation} 
     \Gamma(\omega,\omega',t)=\int_{0}^{t} dt_1 \int_{0}^{t} dt_2 e^{i (\omega t_1 - \omega' t_2)} C(t_{1}-t_{2}),
 \end{equation} 
substituting this, we now have
\begin{align}
\xi(\omega,\omega',t)&=  \frac{1}{2i} \int_{0}^{t} dt_{1}\int_{0}^{t} dt_{2}  (2 \theta(t_{1}-t_{2})-1) e^{i(\omega t_1- \omega' t_2)}   C(t_{1}-t_{2}), \\
&=\frac{1}{i} \int_{0}^{t} dt_{1}\int_{0}^{t} dt_{2}  \theta(t_{1}-t_{2}) e^{i(\omega t_1- \omega' t_2)}   C(t_{1}-t_{2}) -\frac{\Gamma(\omega,\omega',t)}{2i}.
\end{align}
Since the theta forces $t_{1}>t_{2}$ then 
\begin{align}\label{eq:xi_subs}
\xi(\omega,\omega',t)
&=\frac{1}{i} \int_{0}^{t} dt_{1}\int_{0}^{t_{1}} dt_{2}  e^{i(\omega t_1- \omega' t_2)}   C(t_{1}-t_{2}) -\frac{\Gamma(\omega,\omega',t)}{2i},
\end{align}
by substituting Eq. \eqref{eq:gamma_subs} into Eq. \eqref{eq:xi_subs}
\begin{align}
\xi(\omega,\omega',t)
&=\frac{1}{2 i} \left( \int_{0}^{t} dt_{1}\int_{0}^{t_{1}} dt_{2}  e^{i(\omega t_1- \omega' t_2)}   C(t_{1}-t_{2}) -  \int_{0}^{t} dt_2 \int_{0}^{t_{2}} dt_1 e^{i (\omega t_{1} - \omega' t_{2})} C(t_{1}-t_{2}) \right).
\end{align}
At this point notice that this expression contains the same integrals as \eqref{eq:decay_decompose} as a difference instead of a sum, this should not come as a surprise as the decay and Lamb-Shift are associated to the real and imaginary part of the correlation function \cite{breuer,rivas}. After integration the Lamb-Shift can be written as 
\begin{align}\label{eq:cum_diffls}
\xi(\omega,\omega',t) &=  \sum_{k}\xi^{k}(\omega,\omega',t) =\frac{1}{2 i}\sum_{k} (  \chi_{k}(\omega,\omega',t)-\overline{\chi_{k}}(\omega',\omega,t))
\end{align}
we then take the limit $\omega \to \omega'$
\begin{align}\label{eq:cum_same_ls}
\xi^{k}(\omega,\omega,t) &=\lim_{\omega \to \omega'}\xi^{k}(\omega,\omega',t) =  \Im\left(\frac{c_{k}\left((\nu_{k} t - i \omega t- 1 )+e^{-(\nu_{k}-i\omega) t}\right)}{(\nu
_{k}-i\omega)^{2}} \right).
\end{align}
Let us recall that the usual expression used to calculate the Lamb-shift is
\begin{align}\label{eq:shifts}
    &\xi(\omega,\omega',t)=\frac{t^{2}}{2 \pi^{2}} \int_{-\infty}^{\infty} d\phi \sinc\left( \frac{\omega-\phi}{2} t\right) \sinc\left( \frac{\omega'-\phi}{2} t\right) P.V \int_{0}^{\infty} d\nu J(\nu) \left[ \frac{n(\nu)+1}{\phi-\nu} + \frac{n(\nu)}{\phi+\nu} \right].
\end{align}
Compared to the standard way Eqs. \eqref{eq:cum_diffls} and \eqref{eq:cum_same_ls} are remarkably simple. The main advantages of these expressions are two, namely
\begin{itemize}
    \item The same algebraic expressions can be used to obtain both the decay rates and the Lamb-shift term.
    \item The calculation of Lamb-shift is simple, and lifts the requirement of special integration methods for Cauchy's principal value integrals, which are often expensive, noisy and requires special integration methods\cite{li}.
    \item It puts the Lamb-shift corrections and the decay rates back in the same footing, as the imaginary and real parts of the same quantity.
\end{itemize}

\section{Data for the structured Spectral density}\label{app:spectral_density}

The structured spectral density considered corresponds to an experimentally
estimated phonon spectral density of the FMO complex. It was taken from \cite{RatsepJL2007,lorenzoni}. The low-frequency part is modelled by the Adolphs-Renger (AR) spectral density
\begin{equation}
	J_{AR}(\omega)=\frac{S}{s_1+s_2}\sum_{i=1}^{2}\frac{s_i}{7! 2 \omega_i^4}\omega^5 e^{-(\omega/\omega_i)^{1/2}},
\end{equation}
where $S=0.29$, $s_1=0.8$, $s_2=0.5$, $\omega_1=0.069\,{\rm meV}$ and 
$\omega_2=0.24\,{\rm meV}$. The intra-pigment vibrational modes are modelled 
\begin{equation}
    J_{h}(\omega) = \sum_{k=1}^{62} \frac{4 \omega_k s_k \gamma_k (\omega_{k}^2+\gamma_{k}^2)\omega}{\pi((\omega+\omega_k)^{2}+\gamma_{k}^2)((\omega-\omega_k)^{2}+\gamma_{k}^2)},
\end{equation}
where $\omega_k$  is the vibrational frequency and $s_k$  the Huang-Rhys factor. 
whose values are summarized in Table \ref{tab:huangrhys}. The parameter $\gamma_{k}$
is taken to be $ 1\,{\rm ps}^{-1} \approx 5\,{\rm cm}^{-1}$\cite{lorenzoni}. The total phonon 
spectral density of the FMO complex is given by $J(\omega)=J_{\rm AR}(\omega)+J_{h}(\omega)$.
\begin{table}[H]
\caption{Vibrational frequencies $\omega_k$ and Huang-Rhys factors $s_k$ \cite{RatsepJL2007,lorenzoni}}
\label{tab:huangrhys}
\centering
\renewcommand{\arraystretch}{0.75} 
\begin{tabular}{lrrrrrrrrrr}
\toprule
$k$ & 1 & 2 & 3 & 4 & 5 & 6 & 7 & 8 & 9 & 10\\
$\omega_k\,[{\rm cm}^{-1}]$ & 46 & 68 & 117 & 167 & 180 & 191 & 202 & 243 & 263 & 284 \\
$s_k$ & 0.011 & 0.011 & 0.009 & 0.009 & 0.010 & 0.011 & 0.011 & 0.012 & 0.003 & 0.008 \\
\midrule
$k$ & 11 & 12 & 13 & 14 & 15 & 16 & 17 & 18 & 19 & 20 \\
$\omega_k\,[{\rm cm}^{-1}]$ & 291 & 327 & 366 & 385 & 404 & 423 & 440 & 481 & 541 & 568 \\
$s_k$ & 0.008 & 0.003 & 0.006 & 0.002 & 0.002 & 0.002 & 0.001 & 0.002 & 0.004 & 0.007\\
\midrule
$k$ & 21 & 22 & 23 & 24 & 25 & 26 & 27 & 28 & 29 & 30 \\
$\omega_k\,[{\rm cm}^{-1}]$ & 582 & 597 & 630 & 638 & 665 & 684 & 713 & 726 & 731 & 750 \\
$s_k$ & 0.004 & 0.004 & 0.003 & 0.006 & 0.004 & 0.003 & 0.007 & 0.010 & 0.005 & 0.004 \\
\midrule
$k$ & 31 & 32 & 33 & 34 & 35 & 36 & 37 & 38 & 39 & 40\\
$\omega_k\,[{\rm cm}^{-1}]$ & 761 & 770 & 795 & 821 & 856 & 891 & 900 & 924 & 929 & 946 \\
$s_k$ & 0.009 & 0.018 & 0.007 & 0.006 & 0.007 & 0.003 & 0.004 & 0.001 & 0.001 & 0.002\\
\midrule
$k$ & 41 & 42 & 43 & 44 & 45 & 46 & 47 & 48 & 49 & 50 \\
$\omega_k\,[{\rm cm}^{-1}]$ & 966 & 984 & 1004 & 1037 & 1058 & 1094 & 1104 & 1123 & 1130 & 1162 \\
$s_k$ & 0.002 & 0.003 & 0.001 & 0.002 & 0.002 & 0.001 & 0.001 & 0.003 & 0.003 & 0.009 \\
\midrule
$k$ & 51 & 52 & 53 & 54 & 55 & 56 & 57 & 58 & 59 & 60 \\
$\omega_k\,[{\rm cm}^{-1}]$ & 1175 & 1181 & 1201 & 1220 & 1283 & 1292 & 1348 & 1367 & 1386 & 1431 \\
$s_k$ & 0.007 & 0.010 & 0.003 & 0.005 & 0.002 & 0.004 & 0.007 & 0.002 & 0.004 & 0.002 \\
\midrule
$k$ & 61 & 62 \\
$\omega_k\,[{\rm cm}^{-1}]$ & 1503 & 1545 \\
$s_k$ & 0.003 & 0.003 \\
\bottomrule
\end{tabular}
\end{table}
The spectral density is defined in terms of this factors  as
\begin{align}
	J(\omega) =  \sum_{k} \omega_{k}^{2} s_{k} \delta(\omega-\omega_{k}),
\end{align}
We will rescale the coupling strength to the experimentally measured factors via 
a constant $g$ 
\begin{align}
	J_{new}(\omega) = g  \sum_{k} \omega_{k}^{2} s_{k} \delta(\omega-\omega_{k}) = \sum_{k} \omega_{k}^{2} s'_{k} \delta(\omega-\omega_{k}).
\end{align}
For the simulations shown we choose $g=\frac{1}{15}$.
\section{Heat for the Redfield and GKLS master equations}\label{app:current}

Let us remember that the GKLS master equation in the interaction picture is given by
\begin{align}
	\pdv{\rho_{S}^{I}(t)}{t} & =
	\sum_{\omega} \gamma(\omega)  \left( A(\omega)  \rho^{I}_{S}(t) A^{\dagger}(\omega)- \frac{ \{A^{\dagger}(\omega) A(\omega),\rho^{I}_{S}(t)\}}{2} \right) +i \sum_{\omega,} S_{ls}(\omega) \Big[ \rho^{I}_{S}(t), A^{\dagger}(\omega) A(\omega)\Big]                                                                      \\
	                         & = i \sum_{\omega} S_{ls}(\omega) \Big[ \rho^{I}_{S}(t), A^{\dagger}(\omega) A(\omega)\Big] + D_{GKLS}(\rho),
\end{align}
where $D_{GKLS}$ describes the dissipative part of the generator,and $S_{ls}$ is the Lamb-shift coefficient. Now let us notice that

\begin{align}
	\Tr\Big[A[B,C]\Big] & = \Tr\Big[A(BC-CB)\Big] = \Tr\Big[ABC-BAC\Big]  = \Tr\Big[[A,B]C\Big].
\end{align}

using this and Eq. \eqref{eq:heat}, we notice that, the contribution due to the Lamb-shift is proportional to
\begin{align}
	[H_{S},A^{\dagger}(\omega')A(\omega)] & = [H_{S},A^{\dagger}(\omega')]A(\omega) + A^{\dagger}(\omega')[H_{S},A(\omega)] \\
	                                                        & =\omega' A^{\dagger}(\omega')A(\omega) - \omega A^{\dagger}(\omega')A(\omega)   =(\omega'-\omega) A^{\dagger}(\omega')A(\omega),
\end{align}
For the GKLS master equation  $\omega=\omega'$, then we see that only the dissipative part contributes
to heat for a GKLS master equation \cite{Kosloff_2013}. On the other hand notice that for the Redfield equation the Lamb-shift depends on two Bohr frequencies  \eqref{eq:redfield} such that 
the coherent contribution is nonzero in general, thus there is a contribution to the heat current coming from the Lamb-shift, with this in mind let us define 
\begin{align}
    D_{Redfield}(\rho) &= 	\sum_{\omega,\omega'} \Bigg(\tilde \gamma(\omega,\omega',t)  \left( A(\omega')  \rho A^{\dagger}(\omega)- \frac{ \{A^{\dagger}(\omega) A(\omega'),\rho\}}{2} \right)   + i  \tilde S(\omega,\omega',t) \Big[ \rho(t), A^{\dagger}(\omega) A(\omega')\Big] \Bigg).
\end{align}
Then by substituting $\dot{\rho}(t)$ into \eqref{eq:heat}
\begin{align}
	Q(t) & = \Tr\Big[H D_{i}(\rho) \Big].
\end{align}
where $i$ indicates the GKLS or Redfield generator respectively. Each bath has
a dissipator associated to it. To find the individual heat currents we simply
consider the contribution of each to the $k$ environments to $D_{i}=\sum_{k}D^{k}_{i}$ such that
\begin{align}
	J_{k} & = \Tr\Big[H D_{i}^{k}(\rho) \Big].
\end{align}
Now, we would like to remark the two main differences between 
the Redfield and GKLS generator, at least for the computation of 
heat currents 
\begin{itemize}
	\item The Redfield Lamb-Shift does not necessarily commute with the 
	Hamiltonian of the system, while the GKLS one does; this means that
	for the GKLS it is merely an energy shift while for Redfield it can induce 
	non trivial changes. We find in section \ref{sect:example2} that the inclusion of Lamb-shift is necessary to study heat.
	\item The GKLS master equation is time independent, and the generator does 
	not depend on any information about the state. This often leads to 
	non-zero currents at time zero, which should not be the case.
\end{itemize}
The contribution of the Lamb-shift to heat currents remains largely unexplored, perhaps because it's calculation tends to be cumbersome and principal value integrals which are not always straightforward \cite{ss1,ss2,rwc,breuer}, we hope our proposed methodology helps in the widespread inclusion of the Lamb-shift correction. 
\section{Error Bounds}\label{app:bounds}
In this section, we provide error bounds for the approximation's effect on the dynamics of the methods used in this manuscript. First, we examine the impact of errors on the creation of generators in general. Next, we study error bounds on our approximation of decay rates.
\subsection{Bounds for the Cumulant equation}
The Fréchet    derivative of a matrix function is defined as 
\begin{align}
    L(X,\Delta)=f(X+ \Delta)-f(X) + \mathcal{O}(||\Delta||) .
\end{align}
The Fréchet   derivative of the matrix exponential is given by \cite{higham}
\begin{align}
L(\mathcal{K}_{2},\Delta \mathcal{K}) = \int_{0}^{1} e^{(1-s)\mathcal{K}_{2}}\Delta \mathcal{K}e^{s\mathcal{K}_{2}} ds.
\end{align}
We can use the Fréchet   derivative to characterize the deviation due to the approximation of our generator. Taking the norm we have 
\begin{align}
||L(\mathcal{K}_{2},\Delta \mathcal{K})|| &\leq \int_{0}^{1} e^{(1-s)\mathcal||{K}_{2}||}||\Delta \mathcal{K}||e^{s||\mathcal{K}_{2}||} ds, \\&\leq ||\Delta \mathcal{K}|| \int_{0}^{1} ds  e^{||\mathcal{K}_{2}||} ,
\\&\leq ||\Delta \mathcal{K}|| e^{||\mathcal{K}_{2}||} .
\end{align}
Let $\epsilon$ denote relative error, then 
\begin{align}
    ||\Delta\mathcal{K}|| =  \epsilon || \mathcal{K}_{2}||,
\end{align}
thus 
\begin{align}
||L(\mathcal{K}_{2},\Delta \mathcal{K})|| &\leq \epsilon ||\mathcal{K}_{2}|| e^{||\mathcal{K}_{2}||}  .
\end{align}
This bound also applies to the error committed with numerical integration, let us remark that the matrix structure of the generator ends up being more important than the error committed for these bounds.
\subsubsection{Bounds for the Approximation of decay rates and Lamb-shift}
We approximate integrals of the form 
\begin{align}
    I_{1} =\int_{0}^{t} dt_{1} \int_{0}^{t_{1}} dt_{2} e^{i \omega t_{1} -\omega' t_{2}} C(t_{1}-t_{2}).
\end{align}
Let us consider
\begin{align}
    I_{1}-I_{1}^{A} &= \int_{0}^{t} dt_{1} \int_{0}^{t_{1}} dt_{2} e^{i \omega t_{1} -\omega' t_{2}} \Delta C(t_{1}-t_{2})\\
   I_{1} &=\int_{0}^{t} dt_{1} \int_{0}^{t} dt_{2} e^{i \omega t_{1} -\omega' t_{2}} C(t_{1}-t_{2}),
\end{align}
taking the norm, we have 
\begin{align}
    ||I_{1}-I_{1}^{A} || &= ||\int_{0}^{t} dt_{1} \int_{0}^{t_{1}} dt_{2} e^{i \omega t_{1} -\omega' t_{2}} \Delta C(t_{1}-t_{2})|| , \\
    &=||\int_{-\infty}^{\infty}d\nu\int_{0}^{t} dt_{1} \int_{0}^{t_{1}} dt_{2} e^{i\nu(t_{1}-t_{2})}e^{i \omega t_{1} -\omega' t_{2}} \Delta S(\nu)||, \\
    &=||\int_{-\infty}^{\infty}d\nu f(\nu,\omega,\omega') \Delta S(\nu)||. \
\end{align}
If we assume the error on the power spectrum $|\Delta S(\omega)| = \epsilon_{S}$  is homogeneous, then 
\begin{align}
    ||I_{1}-I_{1}^{A} || &= ||\int_{0}^{t} dt_{1} \int_{0}^{t_{1}} dt_{2} e^{i \omega t_{1} -\omega' t_{2}} \Delta C(t_{1}-t_{2})|| , \\
    &=||\int_{-\infty}^{\infty}d\nu\int_{0}^{t} dt_{1} \int_{0}^{t_{1}} dt_{2} e^{i\nu(t_{1}-t_{2})}e^{i \omega t_{1} -\omega' t_{2}} \Delta S(\nu)|| ,\\
    &=||\int_{-\infty}^{\infty}d\nu f(\nu,\omega,\omega') \Delta S(\nu)|| ,\\
    &\leq \epsilon_{S} ||\int_{-\infty}^{\infty}d\nu  f(\nu,\omega,\omega')   ||, \\
    &\leq \epsilon_{S}  \pi \left| t \sinc\left(\frac{(\omega-\omega')}{2}t\right) \right| ,\\
    &\leq \frac{ 2 \epsilon_{S}  \pi }{|\omega-\omega'|}  ,
\end{align}
where we have used
\begin{align}
    f(\nu,\omega,\omega') &= \int_{0}^{t} dt_{1}\int_{0}^{t_{1}} dt_{2} e^{i(\omega+\nu)t_{1}}e^{-i(\omega'+\nu)t_{2}}\\
    &=\frac{1-e^{it(\nu+\omega)}}{(\nu+\omega)(\nu+\omega')} +\frac{e^{it(\omega-\omega')}-1}{(\omega-\omega')(\nu+\omega)},\\
    &=i t \left(\frac{ e^{-i\frac{(\nu+\omega)t}{2}}\sinc\left(\frac{(\nu+\omega)t}{2}\right)}{(\nu+\omega')} -\frac{ e^{-it\frac{(\omega-\omega')}{2}} \sinc\left(\frac{(\omega-\omega')t}{2}\right)}{(\nu+\omega)} \right).  \\ 
   \int_{-\infty}^{\infty}f(\nu,\omega,\omega') d\nu &=  \begin{cases}
        -  i \pi t e^{i\frac{(\omega-\omega')}{2}t} \sinc(\frac{(\omega-\omega')}{2}t) \quad \text{For $\omega\neq\omega'$} \\
        - i\pi t \quad \text{For $\omega=\omega'$}
    \end{cases}. \\ 
    |t \sinc(a t)| &\leq \frac{1}{|a|}.
\end{align}
When $\omega=\omega'$ the limit yields
\begin{align}
    ||I_{1}-I_{1}^{A} || 
    &\leq \epsilon_{S}  \pi t ,
\end{align}
the fact that the bound grows in $t$, when $\omega' =\omega$ is not problematic, as the integral itself grows in $t$ and thus the magnitude of the error tends to stay roughly constant as shown in figures \ref{fig:cum_time} and \ref{fig:time_red}. To see this clearly, consider 
\begin{align}
    I_{1} &= \int_{0}^{t} dt_{1} \int_{0}^{t_{1}} dt_{2} e^{i \omega t_{1} -\omega' t_{2}}  C(t_{1}-t_{2}), \\
    &=\int_{-\infty}^{\infty}d\nu f(\nu,\omega,\omega')  S(\nu) ,\\
    &= t \int_{-\infty}^{\infty}d\nu  \left(\frac{ e^{-i\frac{(\nu+\omega)t}{2}}\sinc\left(\frac{(\nu+\omega)t}{2}\right)}{(\nu+\omega')} -\frac{ e^{-it\frac{(\omega-\omega')}{2}} \sinc\left(\frac{(\omega-\omega')t}{2}\right)}{(\nu+\omega)} \right) S(\nu).
\end{align}
which grows in t, more explicitly in the limit of $\omega=\omega'$
\begin{align}\label{eq:bound_long}
    I_{1} =  t \int_{-\infty}^{\infty}d\nu \left( e^{-i\frac{(\nu+\omega)t}{2}}\sinc\left(\frac{(\nu+\omega)t}{2}\right) -1 \right) \frac{S(\nu)}{(\nu+\omega)} .
\end{align}
Now let us consider the behavior of this integral for small $t$ \footnote{For this discussion to be valid we have assumed that $S(\omega)$ vanishes as $\omega \to \infty$, in other words we have assumed a cutoff frequency $\omega_{c}$ which we can compare with $\frac{1}{t}$}, we may Taylor expand the time dependent part to obtain 
\begin{align}
    I_{1} \approx - t^{2} \int_{-\infty}^{\infty}d\nu \left(\frac{2(\omega+\nu)}{3}t -\frac{i}{2} \right) S(\nu) +\mathcal{O}(t^{4}) ,
\end{align}
so that for short times the magnitude of the real part of the integral grows as $t^{3}$ while the imaginary part grows as  $t^{2}$. On the other hand for later times  the second term of \eqref{eq:bound_long} grows linearly in $t$ while the first one  vanishes for $t\to \infty$ due to the Riemann-Lebesgue lemma provided that $\frac{S(\nu)}{\omega+\nu}$ is a sufficiently well behaved function \cite{Serov2017}. Thus for long times we have 
\begin{align}
    I_{1} &\approx   -t\int_{-\infty}^{\infty}d\nu   \ \frac{S(\nu)}{(\nu+\omega)} . 
\end{align}
This shows a bound growing in $t$ is not a problem as the magnitude of the integral grows in $t$ or faster, and as such, the magnitude of the relative error stays roughly constant in $t$.
\subsection{Bounds for the $TCL$ equation}
The  $TCL$  equation  is a differential equation with a time local generator of the form 
\begin{align}
\dot{\rho}(t) = \mathcal{L}_{TCL}(t)\rho(t),
\end{align}
the error due to approximations in the generator can be written as  
\begin{align}\label{eq:err_diff}
    \dot{\rho}(t)-\dot{\rho}_{\text{approx}}(t) = \Big(\mathcal{L}_{TCL}(t)\rho(t) -\mathcal{L}_{\text{approx}}(t)\rho_{\text{approx}}(t)\Big).
\end{align}
Let $e =\rho(t)-\rho_{\text{approx}}(t)$ and $\Delta \mathcal{L}=\mathcal{L}_{TCL}(t) -\mathcal{L}_{\text{approx}}(t)$ then \eqref{eq:err_diff} can be recasted as 
\begin{align}
    \dot{e} = \mathcal{L}_{TCL}e + \Delta\mathcal{L} \rho_{\text{approx}}. 
\end{align}
If we now take the norm, we have
\begin{align}
    ||\dot{e}|| &= ||\mathcal{L}_{TCL}e + \Delta\mathcal{L} \rho_{\text{approx}}||, \\&\leq ||\mathcal{L}_{TCL}||\ ||e|| + ||\Delta\mathcal{L}||\ || \rho_{\text{approx}}||.
\end{align}
Let $\epsilon$ be the relative error of the approximation of the generator, then 
\begin{align}
    ||\Delta\mathcal{L}|| =  \epsilon || \mathcal{L}_{TCL}||,
\end{align}
which allows us to write 
\begin{align}
    ||\dot{e}||  &\leq ||\mathcal{L}_{TCL}||\ ||e|| + \epsilon ||\mathcal{L}_{TCL}||\ || \rho_{\text{approx}}||.
\end{align}
Let $u=||e||, b= \epsilon ||\mathcal{L}_{TCL}||\ || \rho_{\text{approx}}||, a=||\mathcal{L}_{TCL}||$ then we have 
\begin{align}
    \dot{u}  &\leq a(t) U(t) + b(t),
\end{align}
which is the differential form of Gromwall's inequality \cite{Qin2016}, using Gromwall's results one can then write 
\begin{align}
    u(t)  &\leq U(0)e^{\int_{0}^{t}ds a(s)}  + \int_{0}^{t}ds  b(s) e^{\int_{s}^{t}ds_{1} a(s_{1})}.
\end{align}
Substituting back our definitions results in 
\begin{align}
    ||e(t)|| \leq \epsilon \int_{0}^{s} dse^{\int_{s}^{t}ds_{1}||\mathcal{L}_{TCL}(s_{1})||} ||\mathcal{L}_{TCL}(s)|| \ ||\rho_{\text{approx}}(s)||.
\end{align}
We can recast this in terms of the error of the approximation as
\begin{align}
    ||e(t)|| \leq  \int_{0}^{t}ds e^{\frac{1}{\epsilon}\int_{s}^{t}ds_{1}||\Delta\mathcal{L}(s_{1})||} ||\Delta\mathcal{L}(s)|| .
\end{align}
where we used $||\rho_{\text{approx}}(s)||\leq1$. However, notice this bound is true with respect to errors in the generator regardless of whether we use numerical integration of approximation by sum of exponentials. As $\epsilon$ denotes the maximum error in the calculation of the decay rates and Lamb-shift. 
\subsubsection{Bounds for the decay rates and Lamb-shift}
Higher orders of $TCL$ require nested products of these integrals as
\begin{align} 
I_{n}=&\int_{0}^{t} dt_{1}\int_{0}^{t_{1}} dt_{2}  \dots\int_{0}^{t_{2n-2}} dt_{2n-1} \overbrace{C(t-t_{2})\dots C(t_{1}-t_{2n-1})}^{f(\vec{\Delta t})} \nonumber \\
&\times e^{i\omega_{1}t\pm i\omega_{2}t_{2} \pm \dots\pm i \omega_{n-1}t_{1}\pm i\omega_{n}t_{n-1}}.
\end{align}
To bound them, let us start by looking at the lowest order, namely integrals of the form 
\begin{align}
    I_{1}(t) &= e^{i( \omega_{0}-\omega_{1})t} \int_{0}^{t} ds C(s) e^{i \omega_{1} s} ,\\
    &= e^{i( \omega_{0}-\omega_{1})t} \int_{-\infty}^{\infty} d\nu S(\nu)  \int_{0}^{t} ds e^{i (\omega_{1}+\nu) s}, \\
    &= -i e^{i( \omega_{0}-\omega_{1})t} \int_{-\infty}^{\infty} d\nu S(\nu)  \frac{e^{i(\omega_{1}+\nu)t}-1}{(\omega_{1}+\nu)}.
\end{align}
Then the error would be 
\begin{align}
    |  I_{1}(t) - I_{1}^{A}(t) |
    &= |-i e^{i( \omega_{0}-\omega_{1})t} \int_{-\infty}^{\infty} d\nu   \frac{e^{i(\omega_{1}+\nu)t}-1}{(\omega_{1}+\nu)} \Delta S(\nu)|, \\
    &= | \int_{-\infty}^{\infty} d\nu   \frac{e^{i(\omega_{1}+\nu)t}-1} {(\omega_{1}+\nu)} \Delta S(\nu)|, \\
    &\approx \epsilon_{S}  |\int_{-\infty}^{\infty} d\nu   \frac{e^{i(\omega_{1}+\nu)t}-1} {(\omega_{1}+\nu)} |, \\
    &\approx  \epsilon_{S} | i \pi |, \\
    &\approx\pi \epsilon_{S} \label{eq:tcl_1}.
\end{align}
where we have assumed that the error is homogeneous so we can take it out the integral and $|\Delta S(\nu)| = \epsilon_{S}$. For higher orders we may write the coefficients using an iterative procedure. Consider one of the terms in the time ordered cumulants of Eq. \eqref{eq:tcl_exact_me_expansion} 
\begin{align}
    I_{n}(t) = \int_{0}^{t} dt_{1} \int_{0}^{t_{1}} dt_{2} e^{i\omega_{n}t+\omega_{n+1} t_{1}} I_{n-1}(t_{2}) C(t-t_{1}).
\end{align}
Higher orders require expressing the difference of the product of the original and approximated functions, namely
\begin{align}
    I_{n}(t) -I_{n}^{A}(t) = \int_{0}^{t} dt_{1} \int_{0}^{t_{1}} dt_{2} e^{i\omega_{n}t+\omega_{n+1} t_{1}} \left(I_{n-1}(t_{2}) C(t-t_{1})-I_{n-1}^{A}(t_{2}) C^{A}(t-t_{1}) \right).
\end{align}
Then assuming the absolute error is homogeneous i.e. $|\Delta C(s)|= \epsilon $ we get
\begin{align}
    ||I_{n}(t) -I_{n}^{A}(t)|| &= ||\int_{0}^{t} dt_{1} \int_{0}^{t_{1}} dt_{2} e^{i\omega_{n}t+\omega_{n+1} t_{1}} \left(I_{n-1}(t_{2}) C(t-t_{1})-I_{n-1}^{A}(t_{2}) C^{A}(t-t_{1}) \right)|| \\
    &\leq \int_{0}^{t} dt_{1} \int_{0}^{t_{1}} dt_{2}  ||I_{n-1}(t_{2}) C(t-t_{1})+I_{n-1}^{A}(t_{2}) \left(C(t-t_{1})- C(t-t_{1})-C^{A}(t-t_{1}) \right)|| ,\\
    &\leq \int_{0}^{t} dt_{1} \int_{0}^{t_{1}} dt_{2}  ||\left(I_{n-1}(t_{2})-I_{n-1}^{A}(t_{2}) \right) C(t-t_{1})|| \ + ||I_{n-1}^{A}(t_{2}) \left(C(t-t_{1})- C^{A}(t-t_{1}) \right)||, \\
    &\leq \hspace{-0.6em} \int_{0}^{t}\hspace{-0.8em} dt_{1} \int_{0}^{t_{1}}\hspace{-1em} dt_{2}  ||\left(I_{n-1}(t_{2})-I_{n-1}^{A}(t_{2}) \right)||\ || C(t-t_{1})|| \ + ||I_{n-1}^{A}(t_{2})|| \ ||\left(C(t-t_{1})- C^{A}(t-t_{1}) \right)||, \\
    &\leq \int_{0}^{t} dt_{1} \int_{0}^{t_{1}} dt_{2}  ||\left(I_{n-1}(t_{2})-I_{n-1}^{A}(t_{2}) \right)||\ || C(t-t_{1})|| \ + \epsilon ||I_{n-1}^{A}(t_{2})||, \\
    &\leq \int d\omega \int_{0}^{t} dt_{1} \int_{0}^{t_{1}} dt_{2}  ||\left(I_{n-1}(t_{2})-I_{n-1}^{A}(t_{2}) \right)||\ ||S(\omega)|| \ + \epsilon ||I_{n-1}^{A}(t_{2})|| .
\end{align}
Let us take a look at the second contribution. We have 
\begin{align} 
||I_{n}|| \leq&\int_{0}^{t} dt_{1}\int_{0}^{t_{1}} dt_{2}  \dots\int_{0}^{t_{2n-2}} dt_{2n-1}||f(\vec{\Delta t})
\times e^{i\omega_{1}t\pm i\omega_{2}t_{2} \pm \dots\pm i \omega_{n-1}t_{1}\pm i\omega_{n}t_{n-1}}||.\\
\leq&\int_{0}^{t} dt_{1}\int_{0}^{t_{1}} dt_{2}  \dots\int_{0}^{t_{2n-2}} dt_{2n-1}||f(\vec{\Delta t})||. \\
\leq&\int_{0}^{t} dt_{1}\int_{0}^{t_{1}} dt_{2}  \dots\int_{0}^{t_{2n-2}} dt_{2n-1}||\prod_{i=0}^{n} C(\tau_{i})||
\end{align}
Using the Fourier transform for each $\tau_{i}$ (which are time differences from the elements in $\vec{t}$, the specific differences are not important for the bound) we get
\begin{align} 
||I_{n}||  
\leq&\int_{-\infty}^{\infty}d^{n-1}\omega\int_{0}^{t} dt_{1}\int_{0}^{t_{1}} dt_{2}  \dots\int_{0}^{t_{2n-2}} dt_{2n-1}||\prod_{i=0}^{n} e^{i \omega_{i} \tau_{i}}S(\omega_{i})||.\\
\leq&\int_{-\infty}^{\infty}d^{n-1}\omega\int_{0}^{t} dt_{1}\int_{0}^{t_{1}} dt_{2}  \dots\int_{0}^{t_{2n-2}} dt_{2n-1}\prod_{i=0}^{n} ||S(\omega_{i})||.\\
\leq&\frac{t^{2n-1}}{(2n-1)!}  \left(\int_{-\infty}^{\infty}d\omega||S(\omega)||\right)^{n} .
\end{align}
Substituting we obtain
\begin{align}
    ||I_{n}(t) -I_{n}^{A}(t)||  
    &\leq \int_{0}^{t} dt_{1} \int_{0}^{t_{1}} dt_{2}  ||\left(I_{n-1}(t_{2}) -I_{n-1}^{A}(t_{2}) \right)||\ || C^{A}(t-t_{1})|| \nonumber \\ &+ \epsilon \int_{0}^{t} dt_{1} \int_{0}^{t_{1}} dt_{2} \frac{t^{2n-3}}{(2n-3)!}  \left(\int_{-\infty}^{\infty}d\omega||S(\omega)||\right)^{n-1} ,\\
    &\leq \int d\omega\ || S(\omega)||  \int_{0}^{t} dt_{1} \int_{0}^{t_{1}} dt_{2}  ||\left(I_{n-1}(t_{2}) -I_{n-1}^{A}(t_{2}) \right)|| \nonumber \\ &+ \epsilon  \frac{t^{2n-3}}{(2n-3)!}  \left(\int d\omega||S^{A}(\omega)||\right)^{n-1} .\label{eq:iterative}
\end{align}
We now assume there exists a cutoff frequency $\omega_{c}$ for which $S(|\omega| > \omega_{c})=0$ . Then using Eq. \eqref{eq:AAA_power} we get
\begin{align}
J_{A} &= \int_{-\omega_{c}}^{\omega_{c}} d\omega\ || S^{A}(\omega)|| , \\
&\leq \sum_{k}|c_{k}|\int_{-\omega_{c}}^{\omega_{c}} \frac{d\omega}{\sqrt{(\nu_{k}^{R})^{2} + (\omega - \nu_{K}^{I})^{2}}}, \\ 
&\leq\sum_{k}|c_{k}| \log \left(\frac{\sqrt{(\nu_{k}^{I}+\omega_{c})^2+ (\nu_{k}^{R})^{2}}+\nu_{k}^{I}+\omega_{c}}{\sqrt{(\nu_{k}^{I}-\omega_{c})^2+ (\nu_{k}^{R})^2}+\nu_{k}^{I}-\omega_{c} }\right).
\end{align}
If $\omega_{c}$ is much larger than the exponents ($\omega_{c} \gg \nu_{k}^{R},\nu_{k}^{I}$) of our approximation then
\begin{align}
J_{A}  
&\leq \sum_{k}|c_{k}|\log \left(\frac{\sqrt{(\nu_{k}^{I}+\omega_{c})^2+(\nu_{k}^{R})^{2}}+\nu_{k}^{I}+\omega_{c}}{\sqrt{(\nu_{k}^{I}-\omega_{c})^2+(\nu_{k}^{R})^2}+\nu_{k}^{I}-\omega_{c} }\right) ,\\
&\leq \sum_{k}|c_{k}|\log \left(4\left(\frac{\omega_{c}}{\nu_{k}^{R}}\right)^{2}\right).
\end{align}
Substituting it back into \eqref{eq:iterative}
\begin{align}
    ||I_{n}(t) -I_{n}^{A}(t)||  
    &\leq \int d\omega\ || S(\omega)||  \int_{0}^{t} dt_{1} \int_{0}^{t_{1}} dt_{2}  ||\left(I_{n-1}(t_{2}) -I_{n-1}^{A}(t_{2}) \right)||+ \epsilon  \frac{t^{2n-3}}{(2n-3)!}  J_{A}^{n-1}.
\end{align}
Consider the case $n=2$,  using \eqref{eq:tcl_1}: 
\begin{align}
    ||I_{2}(t) -I_{2}^{A}(t)||  
    &\leq \int d\omega\ || S(\omega)||  \int_{0}^{t} dt_{1} \int_{0}^{t_{1}} dt_{2}  ||\left(I_{1}(t_{2}) -I_{1}^{A}(t_{2}) \right)||+ \epsilon  t  J_{A}, \\
    &\leq \frac{\pi \epsilon_{S} t^{2}}{2} \int d\omega\ || S(\omega)||+ \epsilon  t  J_{A}.
\end{align}
For $n=3$
\begin{align}
    ||I_{n}(t) -I_{n}^{A}(t)||  
    &\leq \int d\omega\ || S(\omega)||  \int_{0}^{t} dt_{1} \int_{0}^{t_{1}} dt_{2}  ||\left(I_{2}(t_{2}) -I_{2}^{A}(t_{2}) \right)||+ \epsilon  \frac{t^3}{3!}  J_{A}^{2}, \\
    &\leq \left(\int d\omega\ || S(\omega)||\right)^{2}  \frac{\pi \epsilon_{S} t^{4}}{4!} +  \left(J_{A} \int d\omega\ || S(\omega)||+J_{A}^{2} \right) \epsilon  \frac{t^3}{3!} . 
\end{align}
In general, we can write this recursively as 
\begin{align}
    ||I_{n}(t) - I_{n}^{A}(t)|| &\leq \left( \int d\omega ||S(\omega)|| \right)^{n-1} \frac{\pi \epsilon_{S} t^{2n-2}}{(2n-2)!}  + \epsilon \frac{t^{2n-3}}{(2n-3)!} \sum_{k=0}^{n-2} \left( \int d\omega ||S(\omega)|| \right)^{k} J_{A}^{n-1-k}.
\end{align}
Using the geometric series, this can be further simplified to 
\begin{align}
    ||I_{n}(t) - I_{n}^{A}(t)|| &\leq \left( \int d\omega || S(\omega)|| \right)^{n-1} \frac{\pi \epsilon_{S} t^{2n-2}}{(2n-2)!}  + \epsilon \frac{t^{2n-3}}{(2n-3)!} \left[ \frac{1 - r^{n-1}}{1- r} \right] J_{A}^{n-1},
\end{align}
where $r= \frac{\int d\omega || S(\omega)|| }{J_{A}}$. If the approximation of the power spectrum is good then $r \approx 1$ and we get
\begin{align}
    ||I_{n}(t) - I_{n}^{A}(t)|| &\leq \left( J_{A}\right)^{n-1}  \left(\frac{\pi \epsilon_{S} t^{2n-2}}{(2n-2)!}  + \epsilon \frac{t^{2n-3}}{(2n-3)!}  (n-1) \right).
\end{align}
Just as in the previous section, the fact that this bond grows in $t$ is related to the fact that the integrals themselves grow in $t$. Notice that the coefficients of $TCL_{4}$ are integrals of the cumulant coefficients times a phase, similarly,  the $TCL_{2}$ coefficients, correspond to their time derivatives \cite{meanforce,intermediate}, thus following similar logic to the previous section the $TCL_{n}$ coefficients grow as $t^{n}$ or higher powers. 
\section{The impact of more exponents}\label{app:onemore}
When using the numerically exact methods, there is a tradeoff between the accuracy of the result and the computational cost associated to it. To show how this works for the approximate methods we describe here, and the HEOM method, we take the system from section \ref{sect:example4}. We choose this system  because in the weak coupling limit the rotating wave approximation is valid and we can have an exact solution without the use of exponents by means of  \eqref{eq:volterra}. We use an Ohmic spectral density with high cutoff, so that convergence to the exact result requires a relatively high number of exponents. 

\begin{figure}[h]
     \begin{subfigure}[b]{0.3\linewidth}
\includegraphics[width=\textwidth]{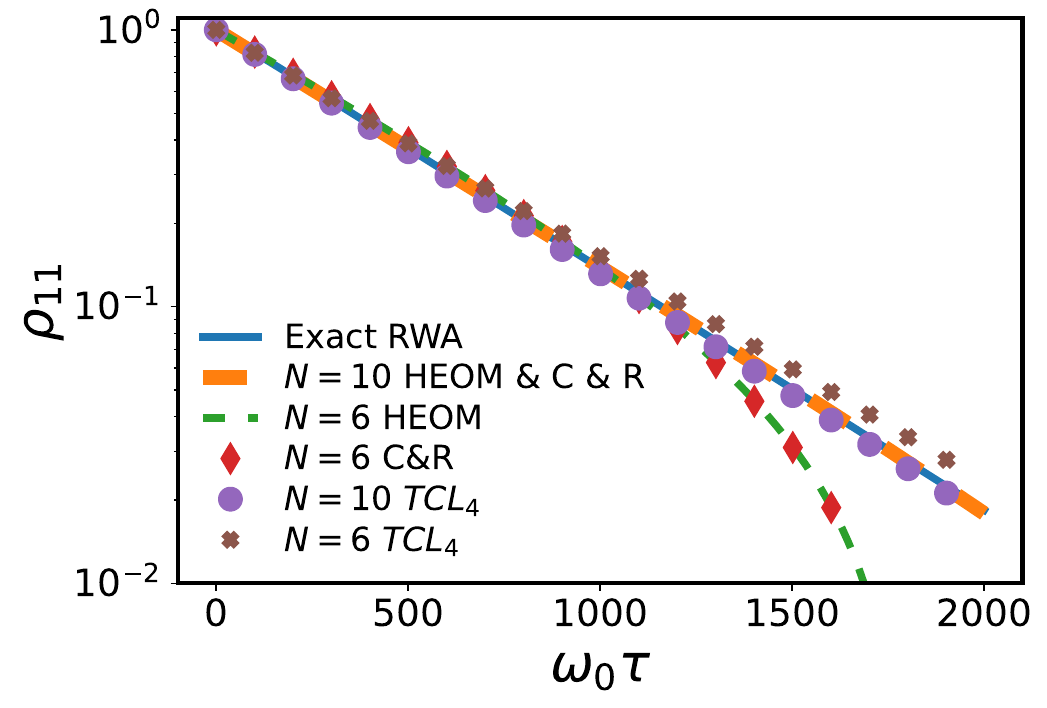}
         \caption{}
         \label{fig:exponents3}
     \end{subfigure}
     \hfill
     \begin{subfigure}[b]{0.3\linewidth}         \includegraphics[width=\textwidth]{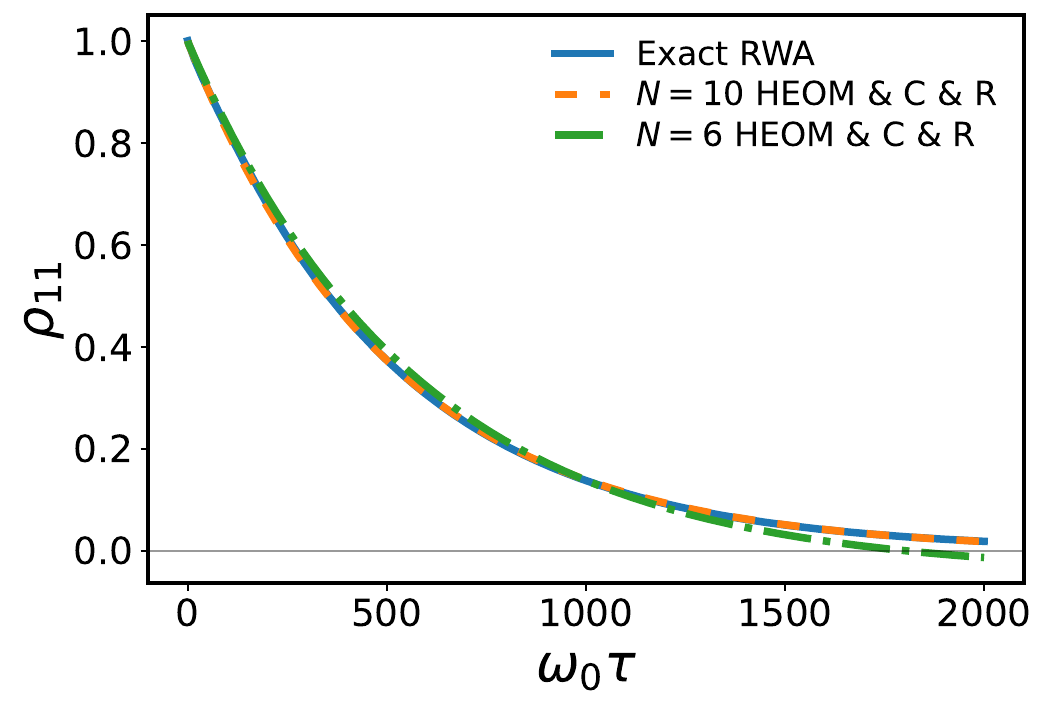}
         \caption{}
         \label{fig:exponents2}
     \end{subfigure}
     \hfill
     \begin{subfigure}[b]{0.3\linewidth}         \includegraphics[width=\textwidth]{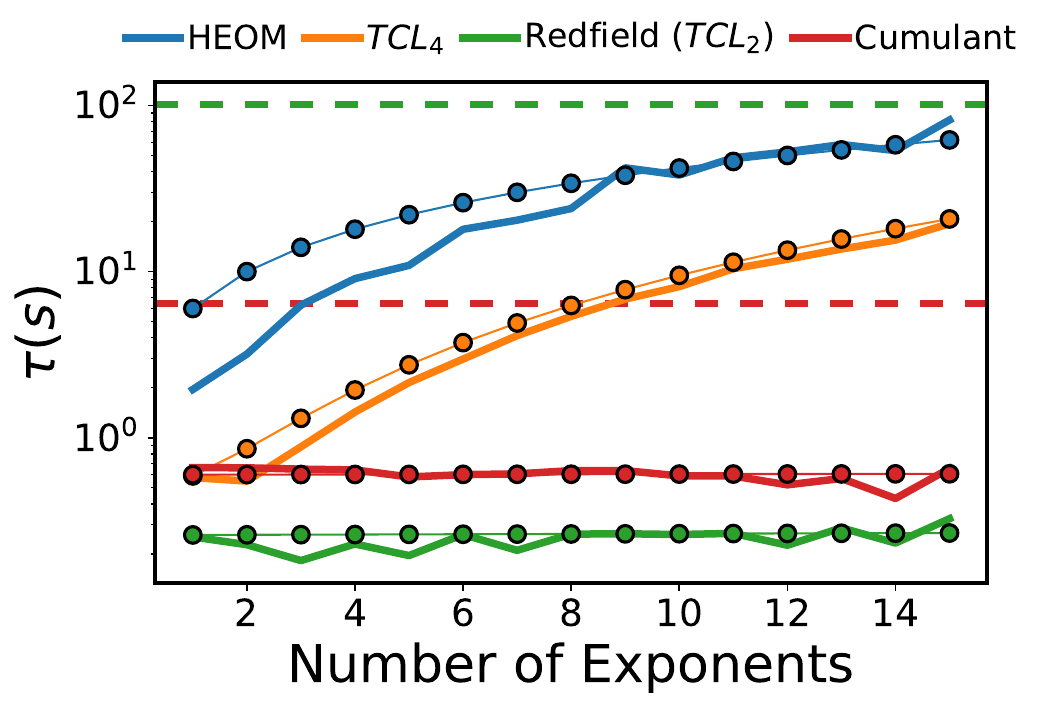}
         \caption{}
         \label{fig:timing2}
     \end{subfigure}
        \caption{Example 6 with an Ohmic spectral density with large cutoff $T=0, \alpha=1 \times 10^{-3}, s=1, \omega_c=80$. (a) The population using the different methods C\&R stands for cumulant and Redfield respectively, while the exact solution corresponds to the volterra equation \eqref{eq:volterra}, the results are shown in the logarithmic scale to highlight their differences (b) Same as (a) but with unit scale on the y axis notice the solution with a small number of exponents goes into negative population and is therefore unphysical (c) The time it takes to run this simulation for different number of exponents, the dashed lines correspond to numerical integration using quadrature methods. While the circular markers correspond to the $\mathcal{O}$ scaling of each method with the number of exponents (linear for cumulant and Redfield, quadratic for $TCL_{4}$ and factorial for HEOM).}
        \label{fig:one_more}
\end{figure}

Another reason to choose this example in this regime is that it preserves the number of excitations, allowing us to know that exact results can be obtained with a HEOM hierarchy of $Nc=1$. By fixing this, our only convergence knob is the number of exponents, which allows us to assess the impact of adding exponents using these different methods. Furthermore, in this example, the simulation time in the case of the methods described in this paper is dominated by the determination of decay rates (the time required to diagonalize the $4\times4$ generator is negligible) and we can see the $\mathcal{O}(M^{n})$ scaling where $M$ is the number of exponents, and compare it with the scaling of the matrix size for HEOM $d_{HEOM}=d^{2}\frac{(M+N_c)!}{( M)!N_c!}$ where $N_c$ is the truncation of the hierarchy and $d$ the dimensionality of the system (here $d=2$). In the case of HEOM, the simulation time will be dominated by the matrix operations. The matrix defines $2 d_{HEOM}$ ODEs.  Solving these ODEs has complexity $\mathcal{O}\left( 4 d_{HEOM}^{2} \right)$. For $N_{c}=1$ our scaling $\mathcal{O}(M^{n})$ is only better if $n\leq2$, however for larger $N_{c}$ HEOM quickly becomes worse and worse, and as shown in the Figures \ref{fig:one_more} $n=2$ is more than enough for this simple toy model. 
\end{document}